\documentclass[preprint,11pt,3p,authoryear]{elsarticle}
\newcommand{\yc}[1]{\textcolor{blue}{}}
\newcommand{\cl}[1]{\textcolor{red}{}}
\usepackage{hyperref}
\usepackage{natbib}
\usepackage{svg}
\usepackage{url}
\usepackage{amssymb}
\usepackage{amsmath}
\usepackage{gensymb}
\usepackage{booktabs}
\usepackage{multirow}
\usepackage{threeparttable}
\usepackage{siunitx}
\usepackage{graphicx}
\usepackage{caption}
\usepackage{array}
\usepackage{subcaption}
\usepackage{cleveref}
\usepackage{amsthm}
\usepackage{geometry}
\usepackage{tabularray}
\usepackage{setspace}
\usepackage[final, defaultcolor=red]{changes} 

\usepackage{lineno}
\begin{document}
\doublespacing
\begin{frontmatter}

\title{\added{Differentiable FEM versus PINN for pavement layer inverse analysis: enforcing physics as hard or soft constraints}}

\author[UMass,KAIST-InnoCORE]{Yongjin Choi\fnref{fn1}}
\author[KAIST-ME]{Hyeonbin Moon\fnref{fn1}}
\author[KAIST-AX,KAIST-ME,KAIST-InnoCORE]{Seunghwa Ryu\corref{cor1}}

\fntext[fn1]{These authors contributed equally to this work.}

\cortext[cor1]{Corresponding author}
\ead{ryush@kaist.ac.kr}

\affiliation[UMass]{organization={Department of Civil and Environmental Engineering, University of Massachusetts Amherst},
            city={Amherst},
            postcode={01003}, 
            state={MA},
            country={USA}}

\affiliation[KAIST-InnoCORE]{organization={InnoCORE PRISM-AI Center, Korea Advanced Institute of Science and Technology (KAIST)},
            city={Daejeon},
            postcode={34141}, 
            country={Republic of Korea}}

\affiliation[KAIST-ME]{organization={Department of Mechanical Engineering, Korea Advanced Institute of Science and Technology (KAIST)},
            city={Daejeon},
            postcode={34141}, 
            country={Republic of Korea}}

\affiliation[KAIST-AX]{organization={Department of AX, Korea Advanced Institute of Science and Technology (KAIST)},
            city={Daejeon},
            country={South Korea}}

\begin{abstract}
\begin{singlespace}
Automatic-differentiation-based inverse analysis methods, including physics-informed neural networks (PINNs) and differentiable programming, have recently shown great promise due to their ability to compute accurate gradients and convergence efficiency.
However, their applicability to falling weight deflectometer (FWD) backcalculation remains unexplored.
This study critically evaluates PINN-based inverse analysis for a multilayer pavement system and investigates differentiable finite element method (DiffFEM) as an alternative based on a synthetic benchmark.
The standard PINN does not recover layer moduli because of the sharp domain discontinuities inherent to layered pavement systems.
Although we use an extended PINN with domain decomposition (XPINN), which shows better performance on discontinuous domains, its performance remains highly sensitive to loss weighting and network architecture, and degrades under measurement noise.
By contrast, DiffFEM consistently achieves more accurate, stable, and computationally efficient inversion results.
\added{These results show and quantify the expected advantage of DiffFEM, which enforces the governing physics as a hard constraint in the solver, over PINN-based approaches, where the governing physics is imposed as a soft constraint through the loss function.}
More broadly, the findings suggest that the choice between PINN- and DiffFEM-based inverse analysis needs careful consideration, with DiffFEM offering practical advantages when an efficient and robust differentiable forward solver is available.
\end{singlespace}
\end{abstract}


\end{frontmatter}


\section{Introduction}
\label{sec:intro}
The falling weight deflectometer (FWD) \citep{smith2017fwd} is a widely used device for non-destructive structural evaluation of pavement systems.
In an FWD test, an impulse load is applied at the pavement surface, and an array of geophones records the resulting deflection basin.
This measured basin is then used to backcalculate layer moduli through an iterative optimization process: a forward model predicts the deflection basin for a candidate set of moduli, and the moduli are updated until the predicted response matches the measurements.

Within the quasi-static, layered elastic framework, early efforts approached this inverse problem using database search optimization, as in MODULUS \citep{RohdeScullion1990_Modulus4}, or sampling-based techniques \citep{zhang2003fwd_genetic}.
These methods, however, either compromise generality or become computationally inefficient as the number of layer moduli increases.
Gradient-based approaches, such as EVERCALC \citep{harichandran1993modified}, provide faster convergence with flexibility in formulation.
Yet, despite their widespread use, these methods share a fundamental numerical limitation: gradients are typically computed via finite-difference (FD) approximations.
From a computational standpoint, this requires at least $N$ additional forward simulations per iteration for $N$ unknown layer moduli, leading to poor scalability and reduced efficiency as the number of layers increases.
In addition, FD gradient approximations introduce truncation errors that are sensitive to the choice of perturbation step size, potentially degrading the quality of the computed gradients \citep{choi2024inverse} and consequently to the stability of the back-analysis.
These limitations naturally motivate more principled and efficient approaches to gradient evaluation.

Automatic differentiation (AD) \citep{margossian2019review_ad,paszke2017automatic} offers a principled alternative by enabling efficient and accurate gradient computation through a single backward pass, thereby avoiding the truncation errors and scalability limitations of FD.
Physics-informed neural networks (PINNs) \citep{raissi2019physics} are a prominent framework built on this capability, in which the governing equations are embedded into the loss function of a neural network.
In inverse analysis, the unknown parameters are treated as trainable variables together with the network weights and are inferred by minimizing a composite loss of governing-equation residuals and data misfit, enabling physics-constrained inversion.
\added{Owing to these advantages, PINNs have been applied to a wide range of scientific and engineering problems \citep{nabian2021efficient,wang2024estimation,huang2025enriched,wang2025adaptive,li2026fundamental}.}
However, their applicability to multilayered pavement inverse analysis for FWD remains largely unexplored.

Despite their promise, PINNs present notable practical challenges.
Their training is often sensitive to hyperparameter choices, including network architecture, learning rate, and the relative weighting between governing-equation residuals and data misfit in the multi-objective loss function \citep{faroughi2024physics,chou2025impact_pinn_weights,kumar2025deep}.
Poor choices of these hyperparameters can lead to non-convergence or suboptimal solutions \citep{kumar2025deep}, while the practical implications of this sensitivity, including computational cost, are not always made fully explicit.
In addition to these general training issues, FWD backcalculation introduces a problem-specific difficulty.
Pavement systems consist of discrete layers with sharp modulus discontinuities at their interfaces, which correspond to high-frequency features in the solution space.
Standard PINNs are known to exhibit spectral bias, that is, a tendency to learn low-frequency components more easily than sharp transitions \citep{naser2025fundamental,peng2025xpinn_multi_layer}.
This limitation may degrade the representation of interlayer behavior and, in turn, the accuracy and robustness of inversion. 
Although Fourier feature embedding can mitigate spectral bias by enhancing high-frequency representation \citep{liu2025leveraging}, they are more effective for smooth multiscale variations than for sharp material discontinuities \citep{fazliani2025enhancing}.
It therefore remains unclear whether PINN-based methods can solve multilayered pavement inverse problems reliably and efficiently.

These challenges highlight a more fundamental issue: while PINNs leverage AD for efficient gradient computation, they rely on a neural network surrogate that only approximately satisfies the governing equations through a loss function and requires problem-specific modifications.
This raises the question of whether the advantages of AD can be retained within a framework that directly enforces the underlying physics.
Differentiable programming \citep{hu2019difftaichi,innes2019differentiable} enables such a framework by applying AD directly to physics-based numerical solvers.
The resulting differentiable solver enforces the governing equations exactly in the forward solve and obtains accurate parameter gradients from a single backward pass.
Differentiable FEM (DiffFEM) is a differentiable solver based on the finite element method (FEM): in FWD backcalculation, it computes the gradients of surface deflections with respect to the layer moduli while satisfying the governing equations at every inversion iteration.
By using a physics-native forward model instead of a neural network surrogate, DiffFEM also avoids the difficulty of representing sharp layer interfaces.
These considerations motivate the investigation of DiffFEM as a physics-consistent alternative for multilayered pavement inverse analysis.

\added{PINN-based forward and inverse analysis has recently been extended to geotechnical engineering \citep{yuan2025physics}, including soil consolidation \citep{bekele2021physics,zhang2022physics,gong2025prediction}, pile--soil interaction \citep{vahab2023application}, elastoplastic footings \citep{chen2025comprehensive}, and subsurface characterization \citep{rasht2022physics}.
In contrast, differentiable solver-based inverse analysis remains largely unexplored in geotechnical and pavement engineering.
Differentiable solvers may provide more accurate and efficient inversion due to their nature of hard physics constraints.
In fact, \citet{wang2025differentiable} developed a differentiable finite element method and reported substantial improvements over PINN-based inverse analyses of engineering structures.
However, they compared their method with previously reported PINN literature, so the discretization, data, and optimization settings were not fully equivalent.
Their formulation also differs from commonly used differentiable solver frameworks (e.g., JAX-FEM \citep{xue2023jax} and DiffMPM \citep{cheng2025structural}) limiting broader comparison.
Thus, systematic and controlled comparisons between PINN- and differentiable solver-based inverse methods remain limited.}

This study critically evaluates the applicability of PINN-based approaches to FWD backcalculation and investigates DiffFEM as a physics-consistent, AD-based alternative, in a controlled setting. 
Specifically, we first examine the performance of a vanilla PINN in recovering pavement moduli in a multilayered system.
Recognizing that the single-network PINN struggles with the multilayered pavement system with modulus discontinuities, we then adopt an extended PINN formulation with domain decomposition (XPINN) \citep{jagtap2020xpinn}, in which each pavement layer is represented by a separate sub-network.
This progression from a standard PINN to a problem-adapted XPINN allows us to evaluate PINN-based inversion under a formulation that is more suitable for the layered structure of pavements.
\added{Finally, we compare these PINN-based approaches with DiffFEM and quantify the expected advantage of DiffFEM across hyperparameter settings and noise levels.}
\added{Based on this quantification, we provide practical guidance for choosing between the two approaches for the inverse analysis in discontinuous multilayered systems.}

\section{FWD backcalculation background}
 
\subsection{Test procedure and inverse problem}
The falling weight deflectometer (FWD) applies a controlled impulse load to the pavement surface and records the resulting deflection basin at an array of geophones positioned at fixed radial offsets from the load center.
Following established practice~\citep{smith2017fwd, VonQuintus2015LTPPModulus}, peak deflection values are used for structural analysis under a quasi-static, layered elastic assumption---a setting shared by many backcalculation software, including EVERCALC~\citep{harichandran1993modified}, and MODULUS~\citep{VonQuintus2015LTPPModulus}.
 
\begin{figure}[!htbp]
    \centering
    \includegraphics[width=0.5\textwidth]{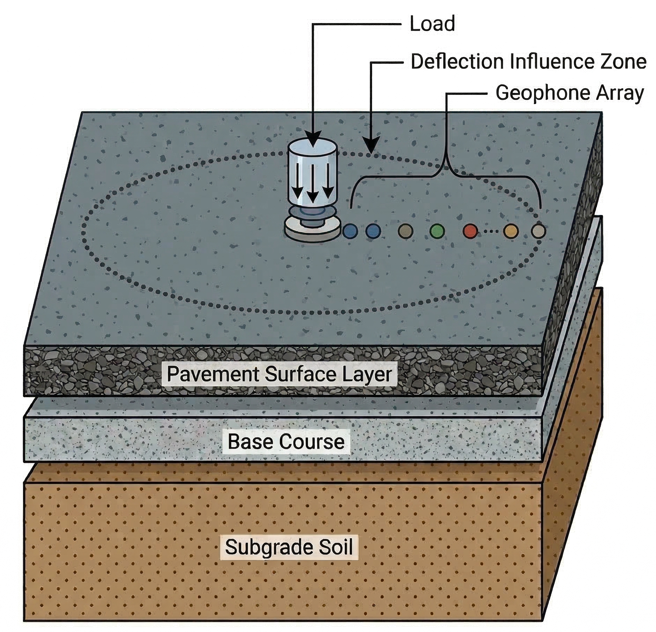}
    \caption{Schematic of the falling weight deflectometer (FWD) test on a layered pavement system.
    An impulse load is applied at the surface, and the resulting deflection basin is recorded by an array of geophones at increasing radial offsets from the load center.}
    \label{fig:overview}
\end{figure}
 
Under Burmister's layered elastic theory~\citep{burmister1945general}, the surface deflections at the $m$ geophone locations are a nonlinear function of the unknown layer moduli $\mathbf{E} = [E_1, \ldots, E_n]^\top$, given known layer thicknesses, Poisson's ratios, and loading conditions (\Cref{eq:forward}). 
 
\begin{equation}
    \mathbf{u} = \mathcal{F}(\mathbf{E}), \qquad
    \mathbf{u} = [u_1, \ldots, u_m]^\top
    \label{eq:forward}
\end{equation}
 
\noindent where $\mathcal{F}: \mathbb{R}^n \to \mathbb{R}^m$ denotes the forward solver that maps the modulus vector to the predicted deflection basin, and $u_j$ is the peak deflection at radial offset $r_j$. 
 
Backcalculation seeks $\mathbf{E}^*$ that minimizes the residual $\ell$ between simulated and measured deflections (\Cref{eq:inverse}).
 
\begin{equation}
    \mathbf{E}^* = \arg\min_{\mathbf{E}}\;
    \mathcal{L}(\mathbf{E}) =
    \frac{1}{m}\sum_{j=1}^{m}
    \ell\!\left(u_j(\mathbf{E}),\, u_j^{\mathrm{obs}}\right)
    \label{eq:inverse}
\end{equation}
 
\noindent where $u_j(\mathbf{E})$ denotes the $j$-th component of $\mathcal{F}(\mathbf{E})$, and $u_j^{\mathrm{obs}}$ is the measured deflection at sensor $j$.
 
Existing approaches address \Cref{eq:inverse} via database search~\citep{RohdeScullion1990_Modulus4}, sampling-based approaches~\citep{gopalakrishnan2010co,zhang2003fwd_genetic}, ANN-based surrogates~\citep{meier1994backcalculation}, or gradient-based iterative solvers~\citep{harichandran1993modified}.
Among them, gradient-based methods exhibit highly efficient convergence behavior.
However, existing implementations share a fundamental numerical limitation regarding scalability and stability, which is discussed next.

\subsection{Finite difference gradient approximation}
\label{subsec:fd}
 
Gradient-based optimization approaches iteratively updates $\mathbf{E}$ based on the gradient information (\Cref{eq:update}) to solve the inverse problem defined in \Cref{eq:inverse}.
\begin{equation}
    \mathbf{E}^{(k+1)} = \mathbf{E}^{(k)}
    - \alpha^{(k)}\,\nabla_{\mathbf{E}}\mathcal{L}(\mathbf{E}^{(k)})
    \label{eq:update}
\end{equation}
\noindent where $\alpha^{(k)}$ is the update step size at iteration $k$.
Each step requires the gradient of the loss with respect to the layer moduli, $\nabla_{\mathbf{E}}\mathcal{L} \in \mathbb{R}^n$.
When $\mathcal{F}$ is evaluated numerically, getting this gradient as a closed-form expression is challenging, so typically estimated from a FD perturbation (\Cref{eq:fd}).

\begin{equation}
    \left[\nabla_{\mathbf{E}}\mathcal{L}\right]_i^{\mathrm{FD}} =
    \frac{\mathcal{L}(\mathbf{E} + \Delta E_i\,\mathbf{e}_i) - \mathcal{L}(\mathbf{E})}
         {\Delta E_i} + \mathcal{O}(\Delta E_i)
    \label{eq:fd}
\end{equation}

\noindent where $\mathbf{e}_i$ is the $i$-th unit vector and $\Delta E_i$ is a user-chosen perturbation applied to modulus $E_i$. 

This approximation has two limitations for multilayer pavement inversion.
First, evaluating the full gradient requires one additional forward solve for each unknown modulus.
For an $n$-layer system, one optimization step therefore requires $n+1$ forward evaluations: one baseline evaluation of $\mathcal{L}(\mathbf{E})$ and $n$ perturbed evaluations.
The per-iteration cost thus scales linearly with the number of unknown moduli, which becomes increasingly expensive as more layers or sublayers are introduced.

Second, the FD gradient is only approximate.
A large perturbation $\Delta E_i$ increases truncation error, whereas a very small perturbation amplifies floating-point cancellation error~\citep{press2007numerical}.
As a result, the gradient quality depends strongly on the perturbation size, and no single choice is uniformly reliable across all layers.
This issue is particularly important in FWD backcalculation, where the surface deflection basin can be only weakly sensitive to some deeper-layer moduli.
In such cases, gradient inaccuracy can misdirect the update in \Cref{eq:update} and destabilize convergence.
 
Automatic differentiation, introduced next, addresses both drawbacks simultaneously.

\subsection{Exact gradients via automatic differentiation}
\label{subsec:ad}

Automatic differentiation (AD) computes derivatives by differentiating the sequence of operations used in the forward computation itself.
Rather than perturbing the input vector $\mathbf{E}$ as in finite differences, AD applies the chain rule to the computational graph of the numerical solver.
Because the derivatives of the elementary operations in that graph are known analytically, AD yields derivatives of the implemented forward model to machine precision, without finite-difference truncation error~\citep{baydin2018automatic}.

For the inverse problem considered here, the quantity of interest is the gradient of the scalar loss $\mathcal{L}$ with respect to the modulus vector $\mathbf{E} \in \mathbb{R}^n$.
Reverse-mode AD is particularly well suited to this setting because one backward pass through the computational graph produces the full gradient with respect to all $n$ unknown moduli:
\begin{equation}
    \nabla_{\mathbf{E}}\mathcal{L}
    \;\xleftarrow{\;\text{1 backward pass}\;}
    \mathcal{L}(\mathcal{F}(\mathbf{E})),
    \label{eq:ad}
\end{equation}

\noindent In contrast to FD, the cost of this reverse pass does not grow linearly with the number of unknown moduli, making AD attractive for inverse problems with multiple parameters.
This shared gradient mechanism underlies both PINN and DiffFEM, whose formulations are described next.

\section{Methodology}
\label{sec:methodology}

\subsection{Differentiable finite element method (DiffFEM)}
\label{subsec:difffem}
 
\begin{figure}[!htbp]
    \centering
    \includegraphics[width=0.75\textwidth]{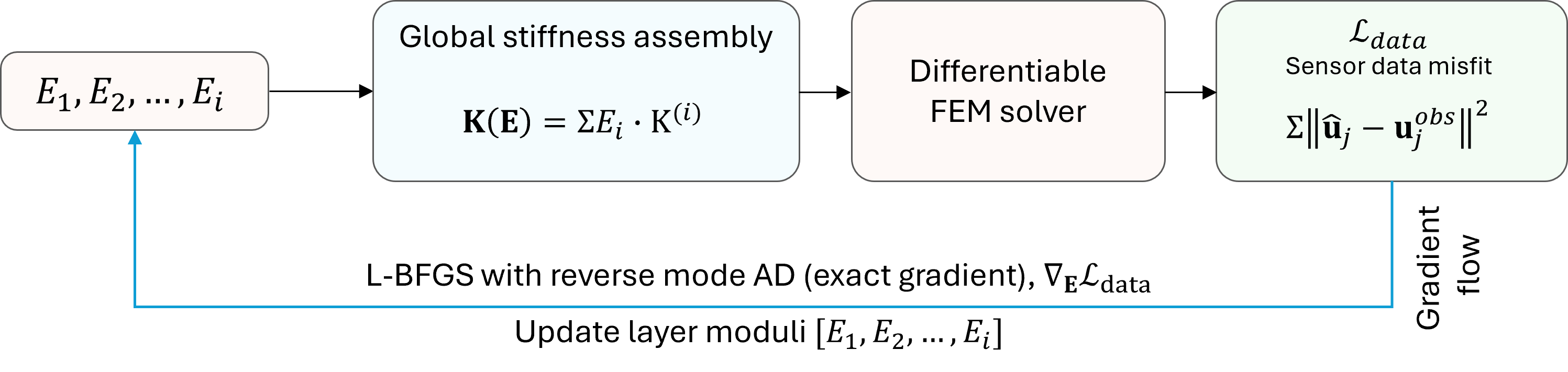}
    \caption{Computational flow of the DiffFEM inversion framework.
    The layer moduli $\mathbf{E}$ enter through the global stiffness assembly, the linear system is solved by a differentiable FEM solver, and the data misfit loss $\mathcal{L}_{\mathrm{data}}$ is evaluated at sensor locations.
    Exact gradients are propagated back to $\mathbf{E}$ via reverse-mode AD (blue arrow).}
    \label{fig:difffem_structure}
\end{figure}

\Cref{fig:difffem_structure} illustrates the DiffFEM framework for pavement layer modulus backcalculation from FWD tests.
The pavement domain is discretized into a two-dimensional axisymmetric finite element mesh using quadrilateral elements, with the radial coordinate $r$ and depth coordinate $z$ as the two spatial dimensions.
For a three-layer system (surface course, base, and subgrade), the global stiffness matrix $\mathbf{K} \in \mathbb{R}^{2N \times 2N}$, where $N$ is the total number of nodes and the factor of two accounts for the radial and vertical displacement degrees of freedom, admits a layer-additive decomposition (\Cref{eq:stiffness_decomp}). 
 
\begin{equation}
    \mathbf{K}(\mathbf{E}) = \sum_{i=1}^{N_{i}} E_i\,\mathbf{K}^{(i)}
    \label{eq:stiffness_decomp}
\end{equation}

\noindent where $N_i$ is the number of layers, $E_i$ is the elastic modulus of layer $i$, and $\mathbf{K}^{(i)}$ is the stiffness matrix assembled from the elements belonging to layer $i$.
 
With boundary conditions applied, the forward problem solves the linear system for the free degrees of freedom (\Cref{eq:ksystem}).
 
\begin{equation}
    \mathbf{K}_{\mathrm{ff}}(\mathbf{E})\,\mathbf{u}_f = \mathbf{F}_f
    \label{eq:ksystem}
\end{equation}
 
\noindent where the subscript $f$ denotes free (unconstrained) degrees of freedom, $\mathbf{u}_f$ is the solution vector of nodal displacements, and $\mathbf{F}_f$ is the corresponding load vector.
All mathematical operations (i.e., the linear combination in \Cref{eq:stiffness_decomp} and the linear solve in \Cref{eq:ksystem}) are recorded in the computational graph through Pytorch autograd \citep{paszke2017automatic}.
The predicted surface deflections at geophone locations are extracted by index-selecting at the corresponding sensor nodes.
 
The inversion seeks the modulus vector $\mathbf{E} = [E_1, E_2, \dots, E_{i}]^\top$ that minimizes the mean squared error between the predicted and observed surface displacements:
 
\begin{equation}
    \mathcal{L}_{\mathrm{data}}(\mathbf{E}) =
    \frac{1}{m}\sum_{j=1}^{m}
    \left\|\hat{\mathbf{u}}_j - \mathbf{u}_j^{\mathrm{obs}}\right\|^2
    \label{eq:loss_diffFEM}
\end{equation}

\noindent where $m$ is the number of sensor nodes, $\hat{\mathbf{u}}_j \in \mathbb{R}^2$ is the predicted displacement vector at sensor node $j$, and $\mathbf{u}_j^{\mathrm{obs}}$ is the corresponding observation.
The gradient $\nabla_{\mathbf{E}}\mathcal{L}_{\mathrm{data}}$ is obtained via a reverse-mode AD backward pass, yielding sensitivities with respect to all moduli.
The resulting gradient is then used by the optimizer to update $\mathbf{E}$. 

\subsection{Physics-Informed Neural Network (PINN)}
\label{subsec:pinn}

\subsubsection{Vanilla PINN}
 
In the vanilla PINN, a single fully connected network $\mathcal{N}_\theta : (r, z) \mapsto (u_r, u_z)$ is trained to predict the displacement field everywhere in the domain.
The network consists of an input layer that maps $(r,z)$ to $H$ neurons, $L$ hidden layers of $H$ neurons each, and a linear output layer (2 neurons), with the hyperbolic tangent activation applied after the input layer and every hidden layer.
The vanilla PINN uses $L=3$ and $H=64$.
 
In standard PINN training, boundary conditions are typically imposed as penalty terms in the loss function \citep{raissi2019physics}, introducing additional weights that complicate loss balancing.
To avoid this issue, we enforce the essential boundary conditions as hard constraints by multiplying the network output by trial functions that vanish at the constrained boundaries \citep{sukumar2022exact}.
This construction satisfies the boundary conditions exactly without an additional penalty term in the loss.
Specifically, the predicted displacements are constructed as:
 
\begin{align}
    \hat{u}_r(r,z) &= r\,(L - r)\,z\;\mathcal{N}_\theta^{(r)}(r,z),
    \label{eq:bc_ur} \\
    \hat{u}_z(r,z) &= z\;\mathcal{N}_\theta^{(z)}(r,z),
    \label{eq:bc_uz}
\end{align}
 
\noindent where $L$ is the domain width, and $\mathcal{N}_\theta^{(r)}$, $\mathcal{N}_\theta^{(z)}$ are the radial and vertical output components of the network, respectively.
This construction exactly satisfies $\hat{u}_r = 0$ at $r = 0$ (axisymmetry), $r = L$ (far-field boundary), and $z = 0$ (rigid base), and $\hat{u}_z = 0$ at $z = 0$, without requiring penalty terms.
 
The total loss combines a physics residual and a data misfit term:
 
\begin{equation}
    \mathcal{L} = \lambda_{\mathrm{phy}}\,\mathcal{L}_{\mathrm{phy}}
                + \lambda_{\mathrm{data}}\,\mathcal{L}_{\mathrm{data}},
    \label{eq:pinn_loss}
\end{equation}
 
\noindent where the physics loss enforces nodal equilibrium using the assembled FEM stiffness matrix:
 
\begin{equation}
    \mathcal{L}_{\mathrm{phy}} =
    \frac{1}{|\mathcal{I}_f|}
    \sum_{i \in \mathcal{I}_f}
    \left\| \bigl[\mathbf{K}(\mathbf{E})\,\hat{\mathbf{u}} - \mathbf{F}
            \bigr]_i \right\|^2,
    \label{eq:loss_physics}
\end{equation}
 
\noindent with $\mathcal{I}_f$ denoting the set of free degree-of-freedom indices.
The data loss $\mathcal{L}_{\mathrm{data}}$ is identical in form to Eq.~\eqref{eq:loss_diffFEM}.
PINN treats both elastic moduli $\mathbf{E}$ and $\theta$ as trainable parameters, so AD computes the $\nabla_\mathbf{E}\mathcal{L}$ and $\nabla_\mathbf{\theta}\mathcal{L}$ and the results are passed to a gradient-based optimization algorithm to jointly updates $\mathbf{E}$ and $\mathbf{\theta}$.

This single network-based PINN architectures typically face convergence challenges for systems with discontinuous domains \citep{jagtap2020xpinn,peng2025xpinn_multi_layer}, such as pavement systems, where the layer's modulus sharply changes.
 
\subsubsection{Extended PINN with domain decomposition (XPINN)}
 
\begin{figure}[!htbp]
    \centering
    \includegraphics[width=1.0\textwidth]{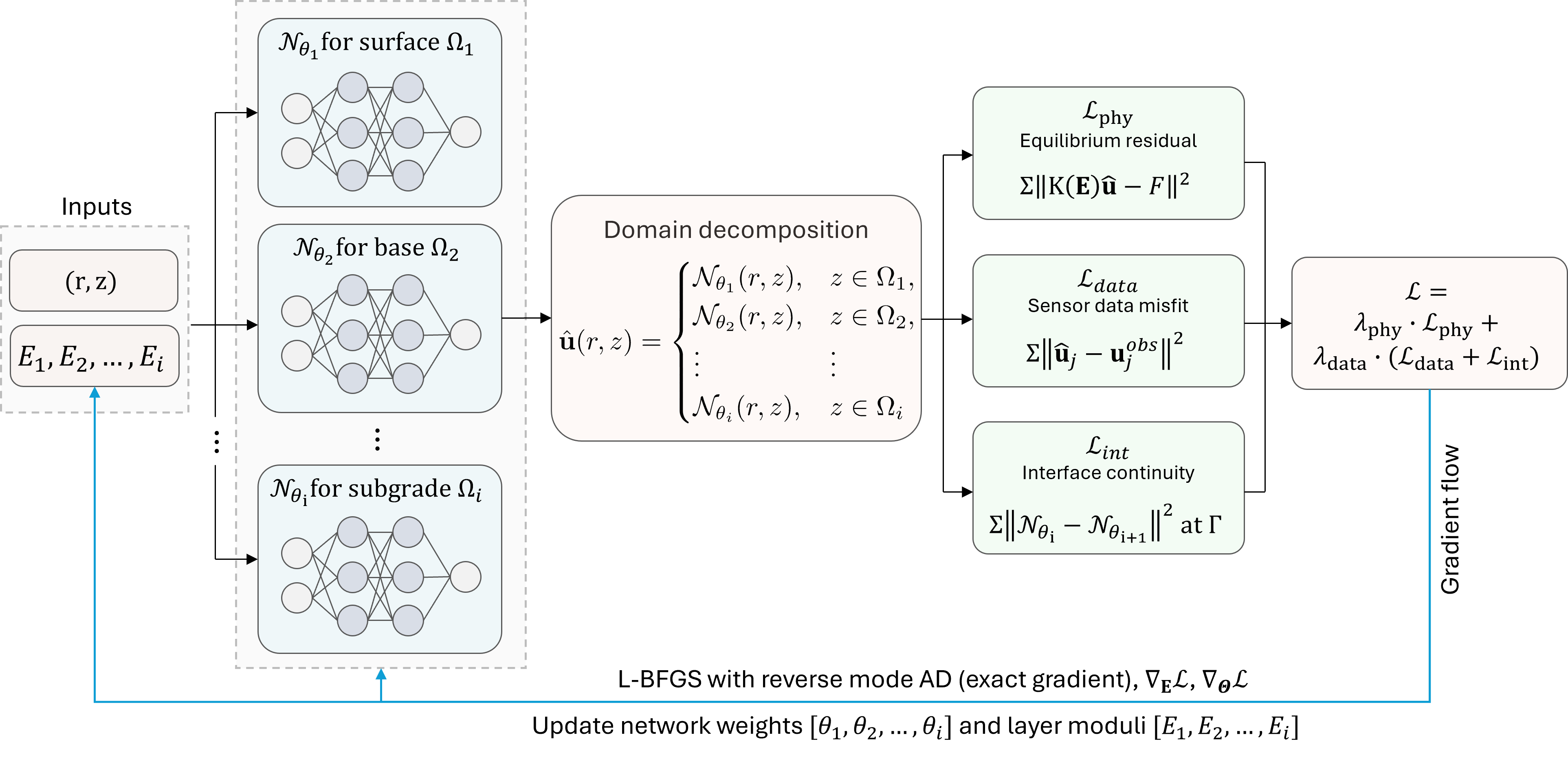}
    \caption{Architecture of the XPINN inversion framework.
    Three independent sub-networks $\mathcal{N}_{\theta_1}$, $\mathcal{N}_{\theta_2}$, $\mathcal{N}_{\theta_3}$ approximate the displacement field in the surface, base, and subgrade domains, respectively.
    The total loss combines a physics residual $\mathcal{L}_{\mathrm{phy}}$, a sensor (i.e., geophone) data misfit $\mathcal{L}_{\mathrm{data}}$, and an interface continuity penalty $\mathcal{L}_{\mathrm{int}}$.
    The layer moduli $\mathbf{E}$ and all network weights are updated jointly via reverse-mode AD (blue arrows).}
    \label{fig:xpinn_structure}
\end{figure}

To address the convergence failure of vanilla PINN under sharp modulus discontinuities at layer interfaces (detailed in Section~\ref{sec:results}), an extended PINN with domain decomposition (XPINN) \citep{jagtap2020xpinn} is adopted in which each pavement layer is assigned a dedicated sub-network.
\Cref{fig:xpinn_structure} shows the overview.
For the three-layer system, three independent networks $\mathcal{N}_{\theta_1}$, $\mathcal{N}_{\theta_2}$, ..., $\mathcal{N}_{\theta_i}$ share the network structure described in the vanilla PINN section, with $L=3$ and $H=32$, corresponding to the surface course, base, and subgrade, respectively.

The composite displacement field is defined piecewise by domain, with each sub-network active exclusively within its assigned layer domain $\Omega_i$.

\begin{equation}
    \hat{\mathbf{u}}(r,z)=
    \begin{cases}
        \mathcal{N}_{\theta_1}(r,z), & z \in \Omega_1,\\
        \mathcal{N}_{\theta_2}(r,z), & z \in \Omega_2,\\
        \vdots & \vdots \\
        \mathcal{N}_{\theta_i}(r,z), & z \in \Omega_{i}
    \end{cases}
    \label{eq:xpinn_composite}
\end{equation}

where $\Omega_i$ denotes the subdomain of layer $i$, $\mathcal{N}_{\theta_i}$ is the neural network associated with that layer $i$.
Boundary conditions are applied to the composite field using the same trial functions as in \Cref{eq:bc_ur}-\eqref{eq:bc_uz}.
 
Displacement continuity at the layer interfaces is enforced as soft constraints appended to the data loss:
 
\begin{equation}
    \mathcal{L}_{\mathrm{int}} =
    \sum_{i=1}^{N_i-1}
    \frac{1}{|\Gamma_{i,i+1}|}
    \sum_{j \in \Gamma_{i,i+1}}
    \bigl\|
        \mathcal{N}_{\theta_i}(\mathbf{x}_j)
        - \mathcal{N}_{\theta_{i+1}}(\mathbf{x}_j)
    \bigr\|^2
    \label{eq:interface_loss}
\end{equation}
 
\noindent where $\Gamma_{i,i+1}$ denotes the set of mesh nodes lying on the interface between layers $i$ and $i+1$.
The data loss in \Cref{eq:pinn_loss} is combined with $\mathcal{L}_{\mathrm{int}}$, so that the total optimization objective becomes:
 
\begin{equation}
    \mathcal{L}_{\mathrm{XPINN}} =
    \lambda_{\mathrm{phy}}\,\mathcal{L}_{\mathrm{phy}} +
    \lambda_{\mathrm{data}}\left(\mathcal{L}_{\mathrm{data}} +
    \mathcal{L}_{\mathrm{int}}\right)
    \label{eq:xpinn_total_loss}
\end{equation}
 
XPINN treats $\mathbf{E}$ and three network parameter sets $\mathbf{\Theta}=\{\theta_1, \theta_2, \ldots, \theta_{i}\}$ as trainable parameters.
Accordingly, AD computes $\nabla_\mathbf{E}\mathcal{L}$ and $\nabla_\mathbf{\mathbf{\Theta}}\mathcal{L}$, and the gradient-based optimization updates the parameters jointly.
Loss weights are set to $\lambda_{\mathrm{phy}} = 1$ and $\lambda_{\mathrm{data}} = 10^3$ following the sensitivity study discussed in \Cref{sec:results}.
 
\subsection{DiffFEM vs. PINN}
 
\Cref{tab:method_comparison} summarizes the structural differences between DiffFEM and XPINN.
The key distinction is that DiffFEM satisfies the governing equations exactly at every iteration: the linear system in \Cref{eq:ksystem} is solved to machine precision.
By contrast, XPINN enforces equilibrium only approximately through the soft physics loss $\mathcal{L}_{\mathrm{phy}}$.
Consequently, DiffFEM optimizes only the three layer moduli, whereas XPINN jointly optimizes the moduli and all network weights across three sub-networks, which total 9,990 parameters (3,330 per sub-network).
 
\begin{table}[htbp]
\centering
\caption{Structural comparison of DiffFEM and XPINN.}
\label{tab:method_comparison}
\begin{tabular}{lll}
\toprule
Item                    & DiffFEM              & XPINN \\
\midrule
Physics enforcement     & Exact (FEM formulation)          & Soft (loss term) \\
Free parameters         & Moduli& Moduli + network weights\\
Interface continuity    & Exact (shared mesh nodes)     & Soft (loss term) \\
Loss terms              & Data only                     & Weighted joint loss with physics + data \\ \bottomrule
\end{tabular}
\end{table}

\subsection{Optimization setting}
 
The gradient-based optimizer used for both DiffFEM and XPINN is the limited-memory Broyden--Fletcher--Goldfarb--Shanno (L-BFGS) algorithm~\citep {liu1989limited}, a quasi-Newton method that uses gradient information to construct a limited-memory approximation of the inverse Hessian and thereby exploit local curvature of the loss function.
We use L-BFGS rather than a first-order optimizer such as Adam since curvature-informed updates can provide faster convergence and lower residuals in PINN-type optimization problems than purely first-order updates~\citep{rathore2024challenges}.
We initialize the moduli to $\mathbf{E} = 1 \ \text{MPa}$, a value deliberately set well below the expected layer stiffnesses to provide an unbiased assessment of convergence behavior for both methods.
Network weights in XPINN are initialized using Xavier uniform initialization~\citep{kumar2017weight}.
\added{We assess convergence with a quantitative criterion common to both methods.}
\added{An iteration denotes one evaluation of the loss and its gradient, including the evaluations within the line search.}
\added{The inversion is regarded as converged at iteration $k$ when, over the preceding $W=\max(20,\lceil 0.05k\rceil)$ iterations, the relative change of every modulus, $|E_i^{(k)}-E_i^{(k-W)}|/|E_i^{(k)}|$, is below $10^{-3}$ and that of the loss, $|\mathcal{L}^{(k)}-\mathcal{L}^{(k-W)}|/|\mathcal{L}^{(k)}|$, is below $10^{-2}$.}
\added{Here, $\mathcal{L}$ is the minimized loss, $\mathcal{L}_{\mathrm{data}}$ (\Cref{eq:loss_diffFEM}) for DiffFEM and $\mathcal{L}_{\mathrm{XPINN}}$ (\Cref{eq:xpinn_total_loss}) for XPINN.}
\added{We evaluate this criterion on the recorded histories, so it does not stop the runs.}
\added{Each run continues until L-BFGS terminates or an iteration budget is reached, $10^{3}$ for DiffFEM and approximately $4\times10^{5}$ for XPINN.}
\added{We report the moduli at the end of each run; where the criterion is met earlier, the moduli at that point differ from the reported values by less than 0.3\%.}

\subsection{Synthetic benchmark setup}
\label{subsec:benchmark}

\begin{figure}[!htbp]
    \centering
    \includegraphics[width=0.5\textwidth]{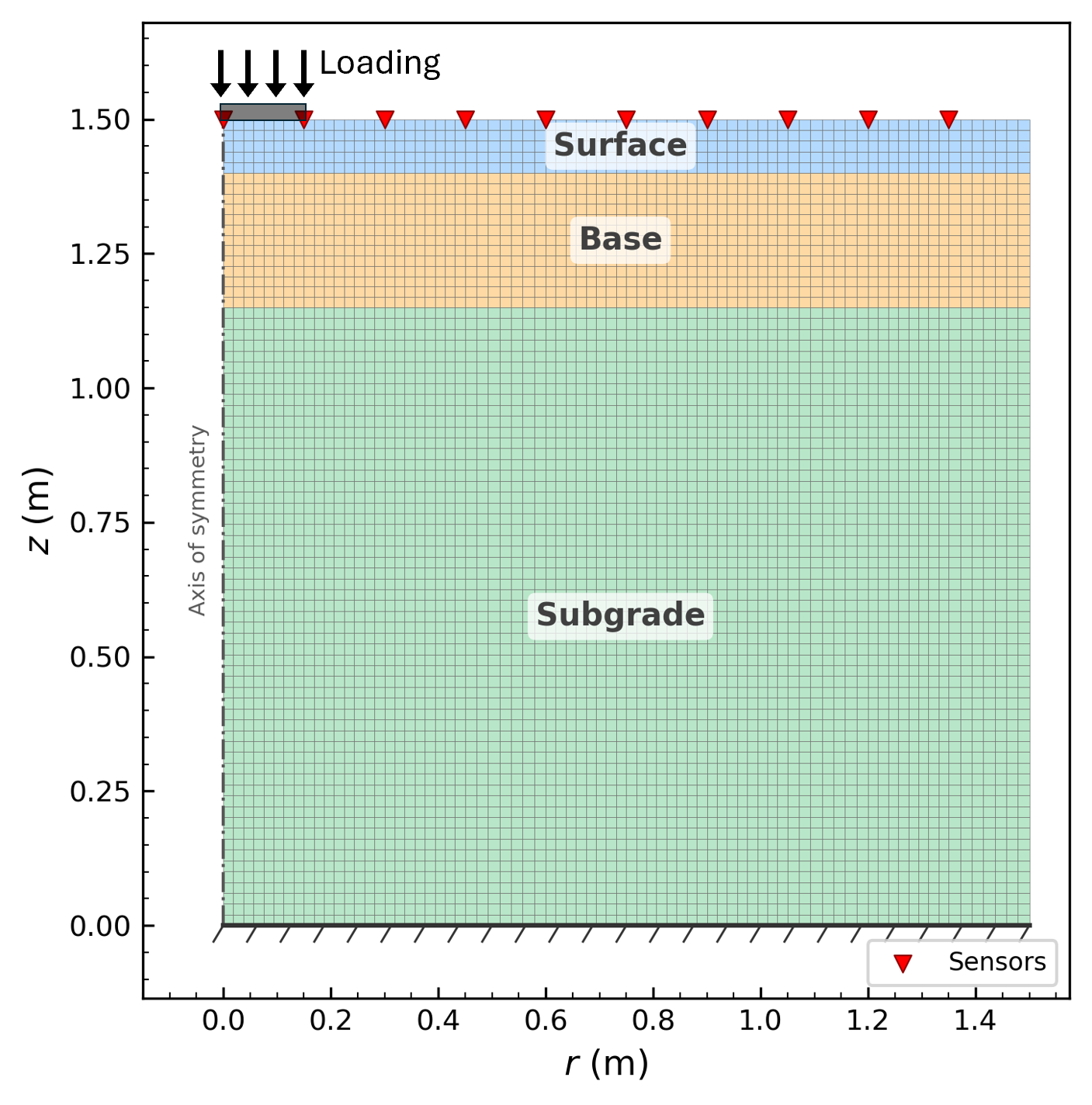}
    \caption{Axisymmetric finite element domain for the three-layer pavement system.
    The mesh spans $r \in [0,\,1.5]$~m and $z \in [0,\,1.5]$~m, with refinement near the loaded zone.
    Ten geophone sensors (red triangles) are placed at the pavement surface at 0.15~m intervals.}
    \label{fig:fe_setting}
\end{figure}

\Cref{fig:fe_setting} illustrates the finite element setup.
A uniformly distributed vertical load is applied over a circular plate of radius $a=0.15$~m at the pavement surface.
The total applied load is approximately 49.5~kN, corresponding to a contact pressure of 0.7~MPa.
The pavement is modeled as a three-layer axisymmetric domain with radial extent $r \in [0,,1.5]$~m and depth $z \in [0,1.5]$~m, where $z=1.5$~m denotes the pavement surface and $z=0$~m denotes the base of the subgrade.
Both the radial and vertical extents are set to five times the loading diameter to minimize boundary effects, following the recommendation of \citep{kuo2004development}.
The layer configuration and reference elastic properties used to generate the synthetic FWD data are listed in \Cref{tab:layers}; these material properties are adopted from \citep{tarefder2014modeling}.
The domain is discretized using 4-node bilinear quadrilateral elements, resulting in a mesh with 78 nodes in the radial direction and 76 nodes in the vertical direction, for a total of 5,928 nodes and 5,775 elements. 
 
\begin{table}[htbp]
\centering
\caption{Layer configuration and reference material properties.}
\label{tab:layers}
\begin{tabular}{llccc}
\toprule
Layer & Depth range & Thickness & $E$ (MPa) & $\nu$ \\
\midrule
Surface (AC)  & $1.40 \leq z \leq 1.50$~m& 100\,mm   & 1{,}378 & 0.35 \\
Base course   & $1.15 \leq z <  1.40$~m& 250\,mm   & 206     & 0.40 \\
Subgrade      & $0.00 \leq z <  1.15$~m& 1{,}150\,mm & 69   & 0.45 \\
\bottomrule
\end{tabular}
\end{table}

\begin{figure}[!htbp]
    \centering
    \includegraphics[width=1.0\textwidth]{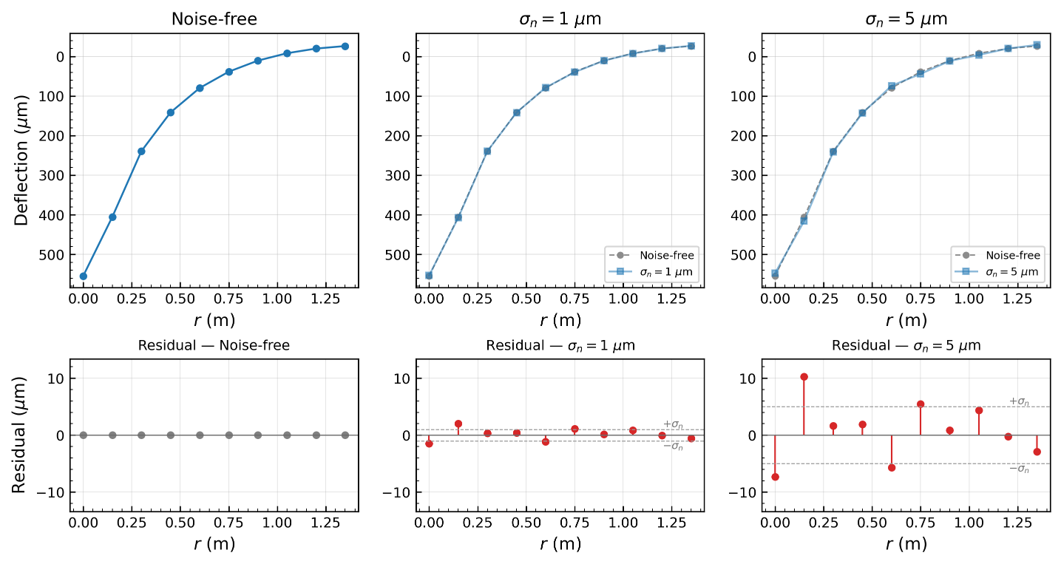}
    \caption{Surface deflection basins measured by FWD geophone sensors under three noise conditions: noise-free, $\sigma_n = 1\,\mu\mathrm{m}$, and $\sigma_n = 5\,\mu\mathrm{m}$.
    Top row: deflection versus radial distance from the load center.
    Bottom row: residuals (noisy $-$ noise-free) at each sensor location; dashed lines indicate the $\pm\sigma_n$ bounds.}
    \label{fig:fwd_basins}
\end{figure}

Ten geophone sensors are located at the pavement surface ($z=1.5$~m) at 0.15~m intervals from $r=0$ to $r=1.35$~m, providing displacement observations $\mathbf{u}$.
To evaluate robustness to measurement uncertainty, six noise cases are considered, corresponding to noise standard deviations of $\sigma_n \in \{0,1,2,3,4,5\}~\mu\mathrm{m}$.
Here, $\sigma_n=0$ represents the noise-free case, while the remaining five cases are generated by adding independent zero-mean Gaussian noise to the displacement at all sensor locations.
\added{This range is selected based on typical FWD measurement errors ($\pm2~\mu\mathrm{m}$) reported by \citet{smith2017fwd}.}
\added{The noise model represents the measurement error of individual geophones.}
For each noise level, the noisy dataset is generated offline prior to inversion and then kept fixed across all methods, so that any performance differences can be attributed solely to the inversion algorithm rather than to stochastic variation in the input data.
\Cref{fig:fwd_basins} illustrates the resulting surface deflection basins for the noise-free case and for two representative noisy cases, $\sigma_n=1~\mu$m and $\sigma_n=5~\mu$m.
The bottom row shows the residuals between noisy and noise-free responses, with dashed lines indicating the $\pm\sigma_n$ bounds.

\added{Because the same finite element model generates the data and performs the inversion, this setting can overestimate the inversion accuracy, a problem known as the inverse crime \citep{kaipio2007statistical}.}
\added{To evaluate the inversion under a discrepancy between the data-generating and inversion models, we also generate the data on a refined mesh that subdivides each element into $2\times2$ elements and add the same noise realizations (\Cref{subsec:model_discrepancy}).}

\section{Results}
\label{sec:results}

This section presents the inversion results for both DiffFEM and XPINN following the synthetic benchmark setup described in \Cref{subsec:benchmark}.

\subsection{Noise-free inversion}
\label{subsec:noise_free}

\begin{figure}[!htbp]
    \centering
    \includegraphics[width=0.75\textwidth]{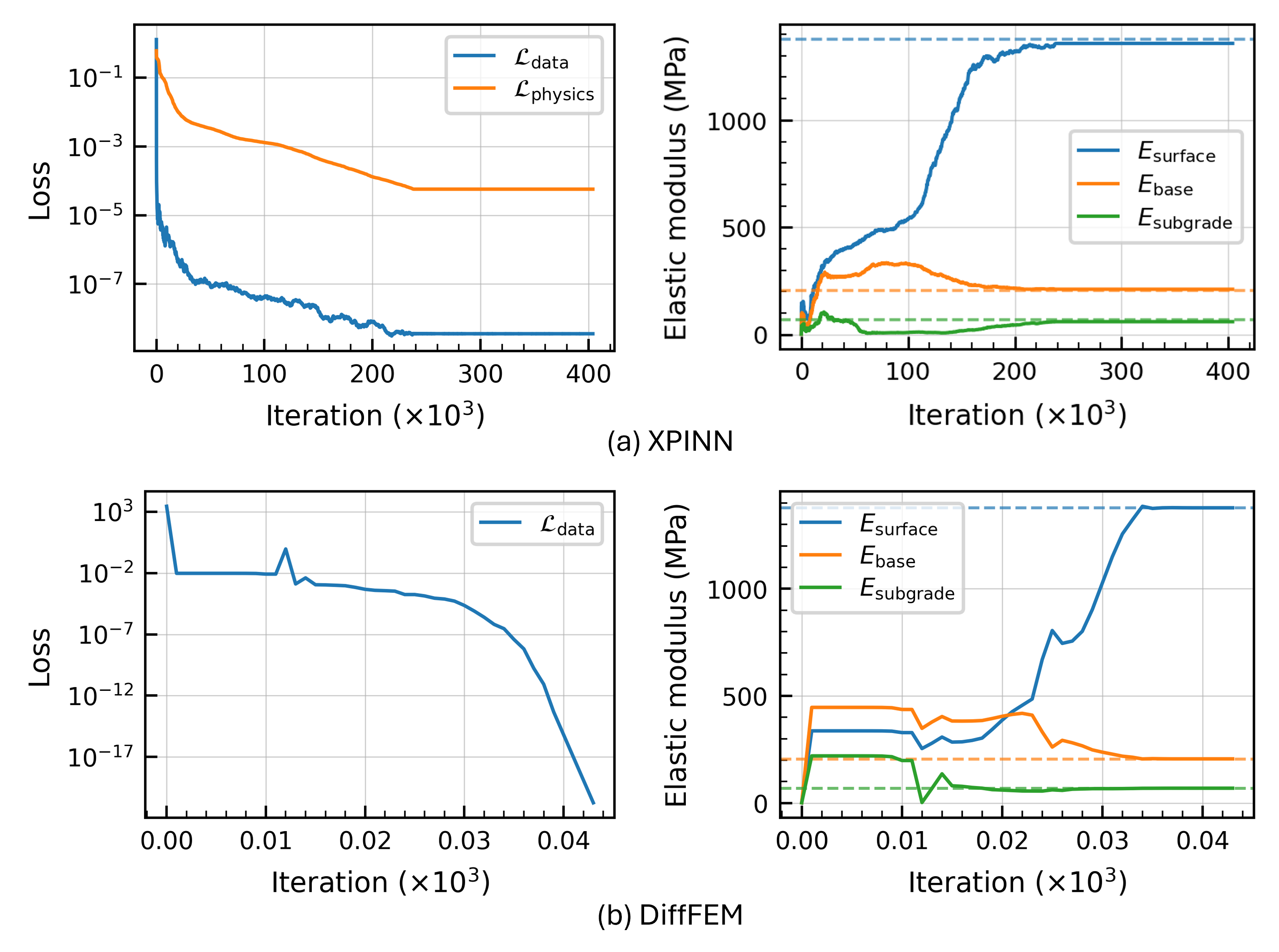}
    \caption{Convergence histories for the noise-free inversion.
    Top row: XPINN loss components (left) and inferred elastic moduli (right) over 400$\times 10^3$ iterations.
    Bottom row: DiffFEM data loss (left) and inferred moduli (right) over fewer than 60 iterations.
    Dashed horizontal lines indicate the ground-truth moduli.}
    \label{fig:inv_hist_noise_0}
\end{figure}

We first evaluate both DiffFEM and XPINN using noise-free measurements to isolate their intrinsic algorithmic behavior from measurement uncertainty.
For XPINN, we adopt loss weights $\lambda_{\mathrm{phy}}=1$, $\lambda_{\mathrm{data}}=10^3$ and a sub-network architecture of $L=3$ hidden layers with $H=32$ neurons, as explained in method section (\Cref{sec:methodology}).
\Cref{fig:inv_hist_noise_0} compares the optimization histories of the two approaches.
DiffFEM (\Cref{fig:inv_hist_noise_0}b) converges within approximately 40 iterations, reducing the data loss below $10^{-17}$ and recovering the three elastic moduli with relative errors on the order of $\mathcal{O}(10^{-9})$–$\mathcal{O}(10^{-10})$ compared to the ground-truth values listed in \Cref{tab:layers}.

In contrast, XPINN (\Cref{fig:inv_hist_noise_0}a) requires roughly $4 \times 10^{5}$ iterations to reach convergence.
Its data loss plateaus near $10^{-8}$, while the physics residual $\mathcal{L}_{\text{phy}}$ stagnates around $10^{-4}$.
Although XPINN estimates moduli that are reasonably close to the true values, the relative errors, -1.5\%, +2.1\%, and -14.4\% for the surface, base, and subgrade layers, respectively, remain several orders of magnitude larger than those obtained with DiffFEM.
This discrepancy can be attributed to the fundamentally different treatments of the governing equations: XPINN enforces the physics as a soft penalty term in the loss function (\Cref{eq:xpinn_total_loss}), which does not strictly guarantee satisfaction of the equations, whereas DiffFEM imposes them as hard constraints through the finite element solver.

\begin{figure}[!htbp]
    \centering
    \includegraphics[width=1.0\textwidth]{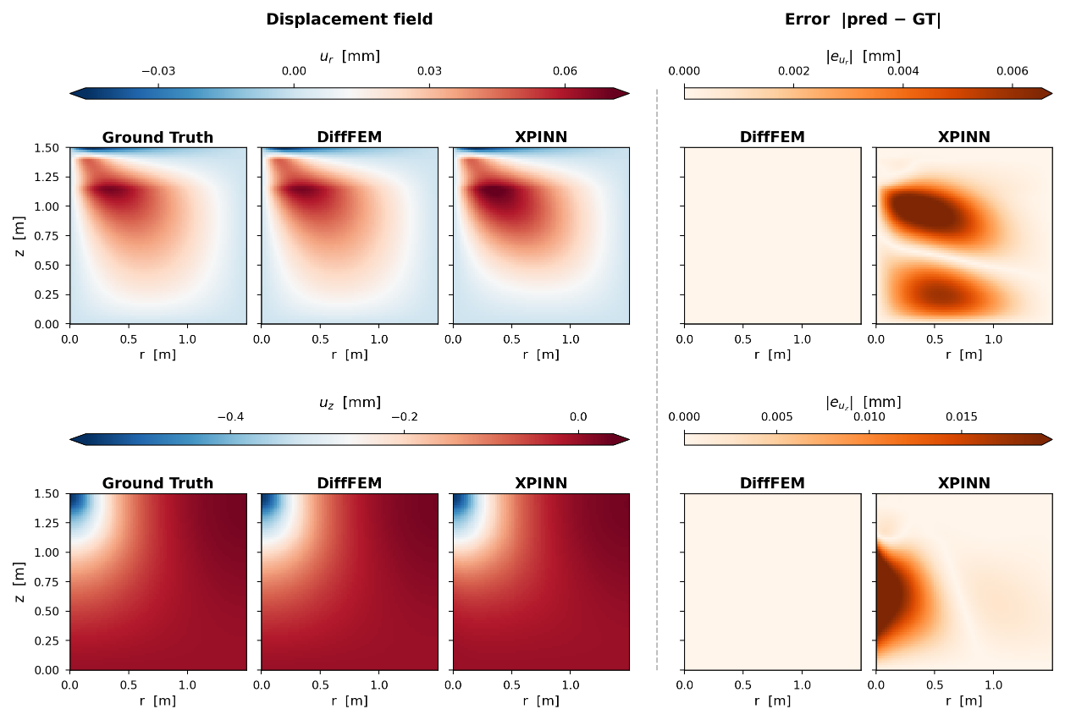}
    \caption{Predicted displacement fields and pointwise errors for the noise-free case.
    Left columns: ground-truth (GT), DiffFEM, and XPINN predictions for radial ($u_r$, top) and vertical ($u_z$, bottom) displacements.
    Right columns: absolute error $|\mathrm{pred} - \mathrm{GT}|$ for DiffFEM and XPINN.}
    \label{fig:displacement_comparison_noise_0}
\end{figure}

These discrepancies in the recovered moduli propagate to the predicted displacement fields, as shown in \Cref{fig:displacement_comparison_noise_0}.
DiffFEM accurately reproduces both $u_r$ and $u_z$ across the entire domain, with negligible error.
XPINN captures the overall displacement patterns but exhibits localized error concentrations: the error in $u_r$ increases below the base-subgrade interface, while the error in $u_z$ is concentrated along the subgrade centerline.

\subsubsection{Limitations and hyperparameter sensitivity of PINN-based inverse analysis}
\label{sec:pinn_limitation_hyperparams}

\begin{figure}[!htbp]
    \centering
    \includegraphics[width=0.4\textwidth]{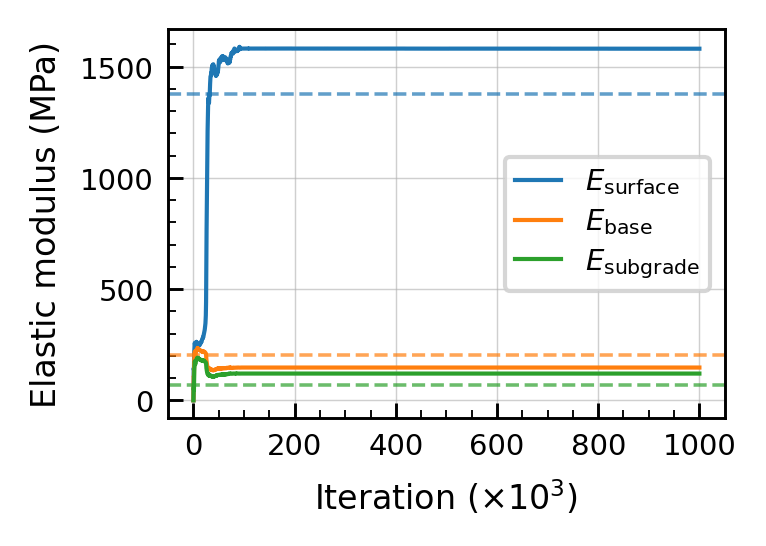}
    \caption{Modulus convergence history for the vanilla PINN formulation without noise.
    Dashed horizontal lines indicate the true moduli.}
    \label{fig:inv_hist_noise_vanilla_pinn}
\end{figure}

The XPINN formulation used in the previous comparison was not the first architecture considered.
As a preliminary study, we also evaluated the vanilla PINN described in \Cref{subsec:pinn}.
However, the vanilla PINN failed to recover the layer moduli, as illustrated in \Cref{fig:inv_hist_noise_vanilla_pinn}.
Specifically, all layer moduli converged to arbitrary values, even in the absence of measurement noise.
This failure is consistent with the well-documented limitations of standard PINNs for discontinuous or multi-domain systems~\citep{jagtap2020xpinn,peng2025xpinn_multi_layer}.
Sharp modulus contrasts at layer interfaces induce slope discontinuities in the displacement field, introducing high-frequency components that a single global network struggles to represent efficiently due to spectral bias~\citep{naser2025fundamental,peng2025xpinn_multi_layer}.
In smooth systems, vanilla PINN may still work well under the noisy conditions \citep{gong2025prediction,moon2025physics}.
However, since our problem, multi-layered pavement systems, has strictly discontinuous modulus changes, the spectral bias issue is expected to be inevitable with standard PINN.
Accordingly, we exclude the vanilla PINN from further consideration and focus the sensitivity analysis in this section on the XPINN-based results.

The composite XPINN objective (\Cref{eq:xpinn_total_loss}) consists of a physics residual term and a data misfit term.
We first investigate the sensitivity to the loss-weight ratio $\lambda_\mathrm{data}/\lambda_\mathrm{phys}$ in recovering the true moduli.
\Cref{tab:xpinn_loss_weight_sweep} summarizes the results of a parameter sweep with the network architecture fixed at $L{=}3$ hidden layers and $H{=}32$ neurons per layer.
When $\lambda_\mathrm{data}/\lambda_\mathrm{phys}=1$, XPINN fails, producing errors exceeding 100\% in all layers.
As the ratio increases, the accuracy improves.
Only when $\lambda_\mathrm{data}/\lambda_\mathrm{phys}=10^3$ do the surface and base moduli converge to within approximately 1-2\% of the ground truth; however, the subgrade modulus still exhibits a substantial relative error of $-14.4\%$.
These results highlight the dependence of PINN performance on loss-weight tuning, a known challenge in PINN-based inverse analysis \citep{chou2025impact_pinn_weights, faroughi2024physics}.
Adaptive loss-weighting schemes \citep{wang2021understanding, mcclenny2023self} aim to mitigate this issue, but they typically introduce algorithmic design choices and computational overhead.

\begin{table}[t]
\centering
\caption{XPINN loss-weight sweep on the noise-free benchmark (network fixed at $L{=}3$ hidden layers, $H{=}32$ units; $\lambda_\mathrm{phys}{=}1$).
Recovered layer moduli (MPa) and signed relative errors (\%) versus the true values listed in \Cref{tab:layers}.}
\label{tab:xpinn_loss_weight_sweep}
    \begin{tabular}{c ccc ccc}
        \toprule
        $\lambda_\mathrm{data}/\lambda_\mathrm{phys}$ & \multicolumn{3}{c}{Recovered modulus (MPa)} & \multicolumn{3}{c}{Relative error (\%)} \\
        \cmidrule(lr){2-4} \cmidrule(lr){5-7}
         & $E_\mathrm{surface}$ & $E_\mathrm{base}$ & $E_\mathrm{subgrade}$ & $E_\mathrm{surface}$ & $E_\mathrm{base}$ & $E_\mathrm{subgrade}$ \\
        \midrule
        $10^{0}$ & -2936 & -44 & -1 & -313.1 & -121.5 & -102.2 \\
        $10^{1}$ & 318 & 289 & 114 & -76.9 & +40.2 & +65.5 \\
        $10^{2}$ & 1454 & 188 & 81 & +5.5 & -8.6 & +17.2 \\
        $10^{3}$ & 1358 & 210 & 59 & -1.5 & +2.1 & -14.4 \\
        \bottomrule
    \end{tabular}
\end{table}

Even after selecting the best-performing loss weights, a second challenge emerges: sensitivity to network architecture.
\Cref{tab:xpinn_network_sweep} presents results obtained using the optimal loss-weight ratio ($\lambda_\mathrm{data}/\lambda_\mathrm{phys}=10^3$) across various architectures with hidden-layer counts $L\in\{2,3,4\}$ and widths $H\in\{16,32,64,128\}$.
The surface modulus is relatively robust, with most configurations yielding errors within a few percent.
The base modulus shows moderate sensitivity, with errors typically ranging from 1\% to 17\%.
In contrast, the subgrade modulus exhibits substantial variability, ranging from $-68.8\%$ at $(L{=}4, H{=}16)$ to $+70.9\%$ at $(L{=}3, H{=}16)$, without a clear trend with respect to network depth or width.
Notably, this architectural sensitivity underscores that the computational effort required for hyperparameter tuning is a prerequisite for achieving PINN's successful convergence.
\added{Because no configuration performs consistently best, the noisy-data experiments below retain $L=3$ and $H=32$.}

\begin{table}[t]
\centering
\caption{XPINN network-size sweep on the noise-free benchmark (loss weights fixed at $\lambda_\mathrm{data}{=}10^{3}$, $\lambda_\mathrm{phys}{=}1$).
Recovered layer moduli (MPa) and signed relative errors (\%) for hidden-layer counts $L\in\{2,3,4\}$ and per-layer widths $H\in\{16,32,64,128\}$, against true values listed in  \Cref{tab:layers}.}

\label{tab:xpinn_network_sweep}
    \begin{tabular}{cc ccc ccc}
        \toprule
        $L$ & $H$ & \multicolumn{3}{c}{Recovered modulus (MPa)} & \multicolumn{3}{c}{Relative error (\%)} \\
        \cmidrule(lr){3-5} \cmidrule(lr){6-8}
         &  & $E_\mathrm{surface}$ & $E_\mathrm{base}$ & $E_\mathrm{subgrade}$ & $E_\mathrm{surface}$ & $E_\mathrm{base}$ & $E_\mathrm{subgrade}$ \\
        \midrule
        2 & 16 & 1303 & 211 & 61 & -5.5 & +2.5 & -12.0 \\
         & 32 & 1374 & 202 & 71 & -0.3 & -1.9 & +3.4 \\
         & 64 & 1254 & 221 & 40 & -9.0 & +7.4 & -42.6 \\
         & 128 & 1361 & 208 & 63 & -1.2 & +0.8 & -8.2 \\
        \midrule
        3 & 16 & 1499 & 173 & 118 & +8.8 & -16.2 & +70.9 \\
         & 32 & 1358 & 210 & 59 & -1.5 & +2.1 & -14.4 \\
         & 64 & 1363 & 208 & 64 & -1.1 & +1.0 & -6.9 \\
         & 128 & 1385 & 202 & 69 & +0.5 & -1.9 & -0.6 \\
        \midrule
        4 & 16 & 1541 & 171 & 22 & +11.8 & -17.1 & -68.8 \\
         & 32 & 1368 & 208 & 61 & -0.8 & +1.2 & -12.3 \\
         & 64 & 1365 & 207 & 64 & -0.9 & +0.3 & -6.8 \\
         & 128 & 1375 & 207 & 62 & -0.3 & +0.5 & -10.6 \\
        \bottomrule
    \end{tabular}
\end{table}

\begin{figure}[!tbp]
    \centering
    \includegraphics[width=0.5\textwidth]{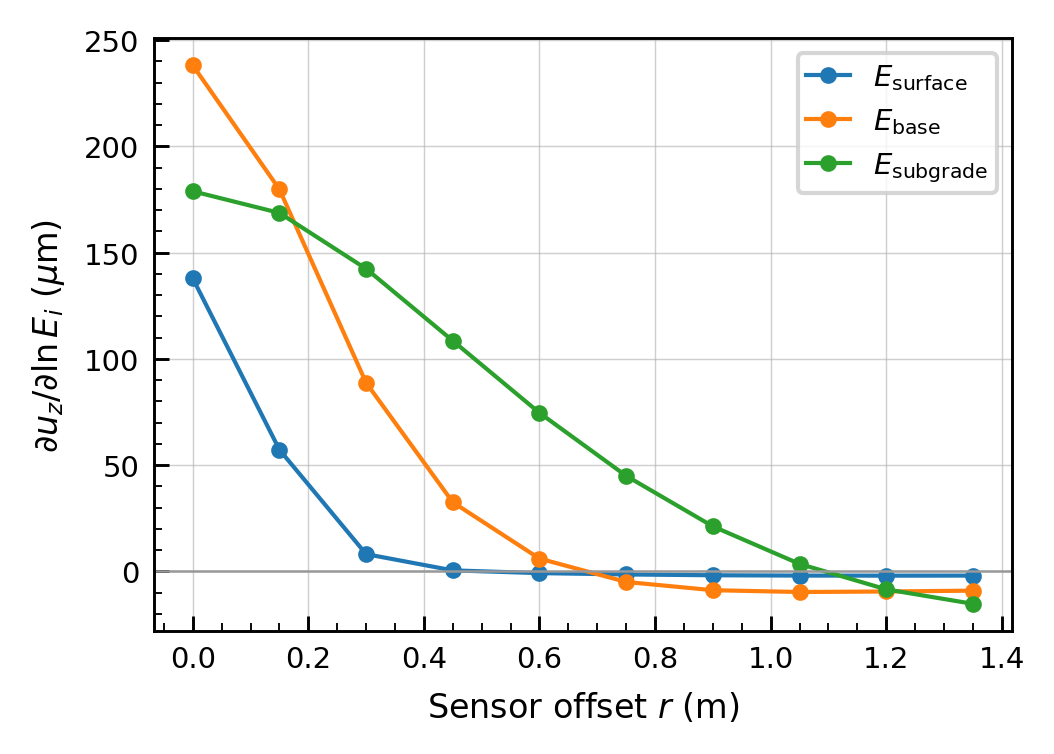}
    \caption{\added{Sensitivity of the vertical surface displacement to each layer modulus at the ten sensors, evaluated at the true moduli.
    A value of 100~$\mu$m means that a 1\% increase in $E_i$ reduces the downward deflection by about 1~$\mu$m.}}
    \label{fig:sensitivity}
\end{figure}

\added{With the selected loss weights, the subgrade error ($-14.4\%$) is much larger than the surface and base errors ($-1.5\%$ and $+2.1\%$), and it varies the most across network architectures.}
\added{To determine whether these subgrade errors stem from the inverse problem itself or from the inversion algorithm, we examine how the surface deflections respond to each layer modulus.}
\added{Because FWD backcalculation relies primarily on vertical deflections, we focus on the vertical displacement; \Cref{fig:sensitivity} shows its sensitivity at each sensor, $\partial u_z/\partial \ln E_i$, computed from the finite element model at the true moduli.}
\added{The surface modulus affects only the two sensors nearest the load, whereas the subgrade modulus has the largest sensitivity from $r=0.3$ to $0.9$~m.}
\added{The deflections therefore constrain the subgrade modulus well, and DiffFEM recovers it exactly from the same data, but XPINN fails.}
\added{The large subgrade errors of XPINN thus arise from the soft-constrained optimization rather than from a weak sensitivity of the data.}

Overall, these results demonstrate that XPINN-based inverse analysis requires careful and problem-specific tuning of both loss weights and network architecture.
Even under noise-free conditions, the method exhibits significant variability in performance, deeper layers as well, which complicates its practical deployment.

\subsection{Noisy inversion}
\label{subsec:noisy}

We next evaluate both methods in the presence of measurement noise, using the same hyperparameters as in the noise-free case for both DiffFEM and XPINN.
The cases of $\sigma_n = 1\,\mu\mathrm{m}$ and $\sigma_n = 5\,\mu\mathrm{m}$ are examined in detail, while the full set of results is summarized in \Cref{subsec:noise_summary} with their quantitative values shown in \Cref{tab:appndx_inversion_summary} in Appendix section.

\subsubsection{Low noise ($\sigma_n = 1 \ \mu\mathrm{m}$)}

\begin{figure}[!htbp]
    \centering
    \includegraphics[width=0.75\textwidth]{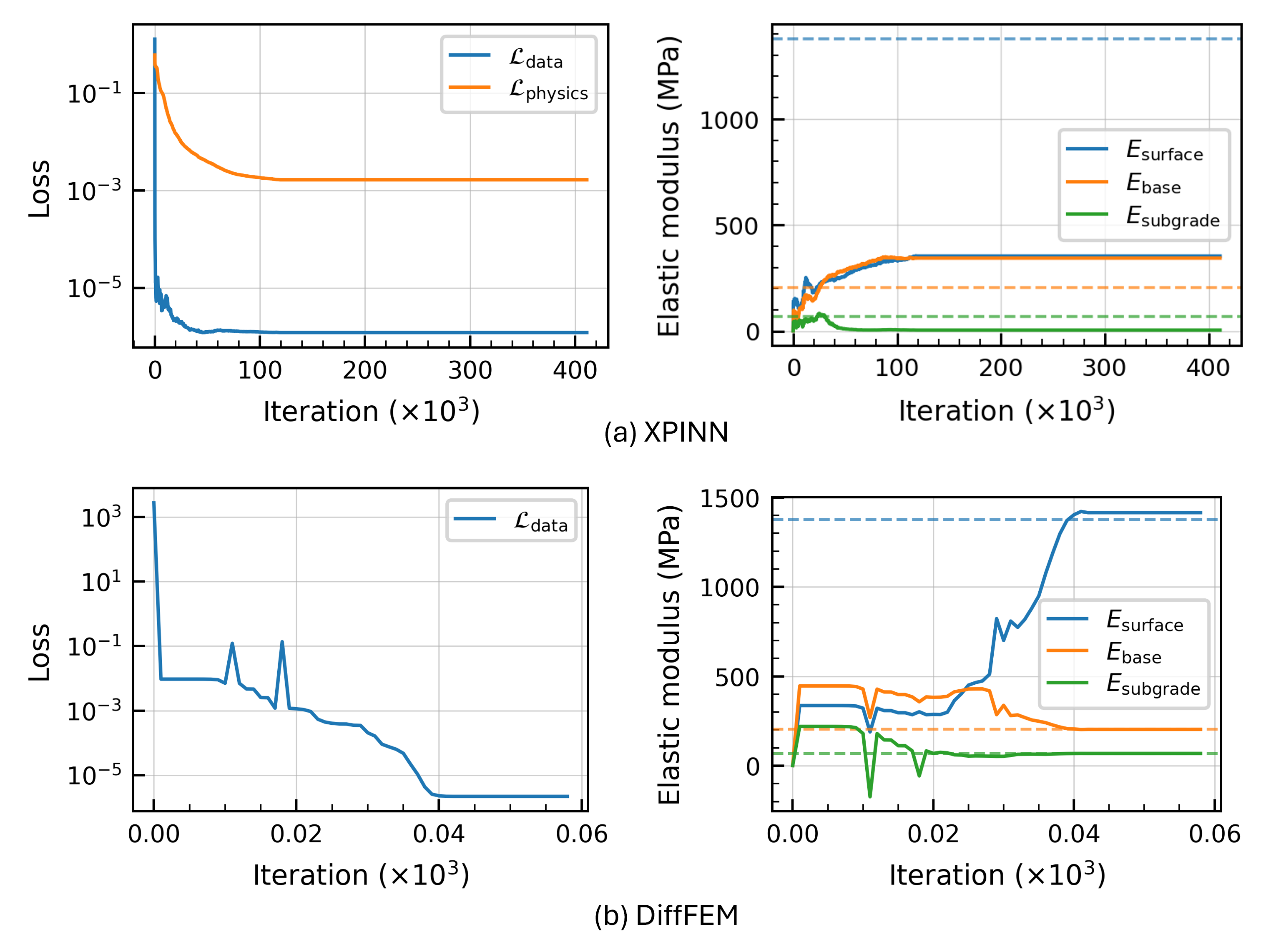}
    \caption{Convergence histories for the noisy inversion with $\sigma_n = 1\,\mu\mathrm{m}$.
    Top row: XPINN loss components and inferred moduli.
    Bottom row: DiffFEM loss and inferred moduli.
    Dashed horizontal lines indicate the ground-truth moduli.}
    \label{fig:inv_hist_noise_1}
\end{figure}

\Cref{fig:inv_hist_noise_1} presents the optimization histories for the case $\sigma_n = 1\,\mu\mathrm{m}$.
DiffFEM (\Cref{fig:inv_hist_noise_1}b) again converges rapidly, after approximately 40 iterations.
The data loss plateaus below $10^{-5}$, and the recovered moduli remain within a few percent of the true values.
Specifically, the relative errors are 2.7\%, 1.2\%, and 0.05\% for the surface, base, and subgrade layers, respectively.
\added{Among the three layers, the surface modulus exhibits the largest relative error, consistent with its sensitivity being confined to the sensors nearest the load (\Cref{fig:sensitivity}).}

\begin{figure}[!htbp]
    \centering
    \includegraphics[width=1.0\textwidth]{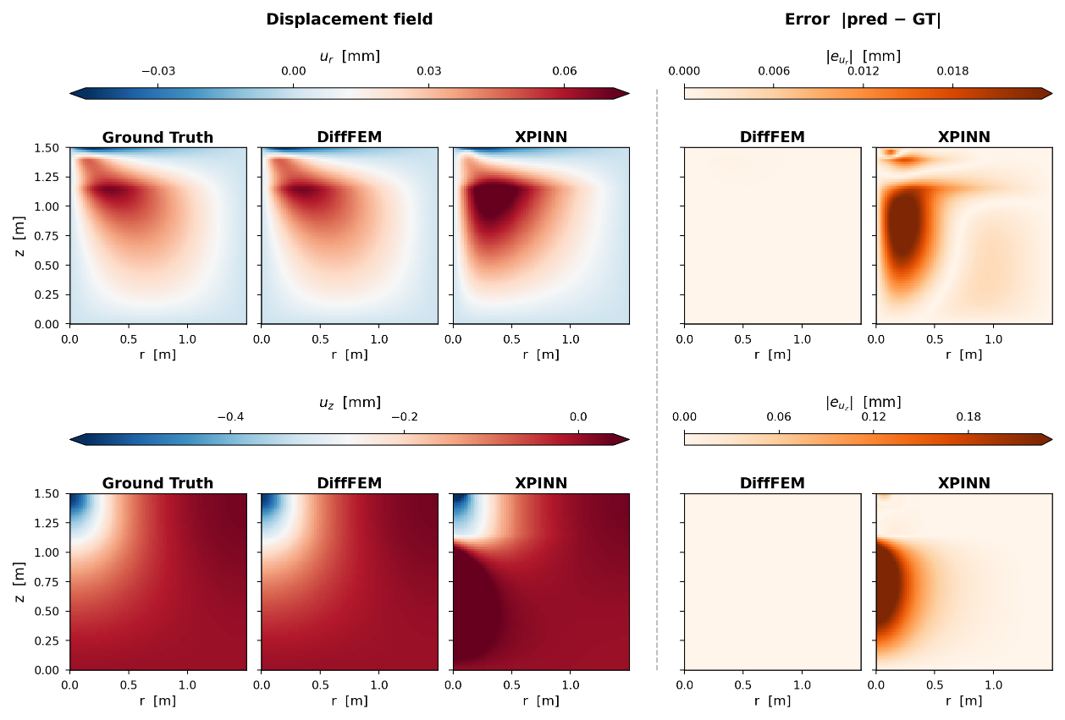}
    \caption{Predicted displacement fields and pointwise errors for $\sigma_n = 1\,\mu\mathrm{m}$.
    The plot layout follows that of \Cref{fig:displacement_comparison_noise_0}.}
    \label{fig:displacement_comparison_noise_1}
\end{figure}

In contrast, XPINN (\Cref{fig:inv_hist_noise_1}a) fails to recover the correct moduli.
Although the data loss reaches a very low plateau (approximately $10^{-6}$), the physics residual stagnates at approximately $10^{-3}$.
The estimated surface and base moduli converge to nearly identical wrong values of about 350~MPa, while the subgrade modulus collapses toward zero, resulting in relative errors of 74\%, 67\%, and 94\%, respectively.
The combination of low data loss and non-negligible physics residual indicates that the recovered displacement field fits well with the sensor observations without satisfying global equilibrium---a solution admissible only under the soft enforcement of the governing equations, which is what PINNs do.
The resulting displacement and error fields in \Cref{fig:displacement_comparison_noise_1} corroborate this interpretation, showing substantial deviations from the ground-truth displacement field throughout the domain.
This behavior suggests that the XPINN failure is not simply a noise-induced perturbation of the inverse solution, but an optimization failure in which the network finds a low data loss displacement field that is not consistent with the correct material parameters.
On the other hand, DiffFEM shows very accurate displacement fields \Cref{fig:displacement_comparison_noise_1} due to the successful modulus recovery.

\subsubsection{High noise ($\sigma_n = 5~\mu \mathrm{m}$)}

\begin{figure}[!htbp]
    \centering
    \includegraphics[width=0.75\textwidth]{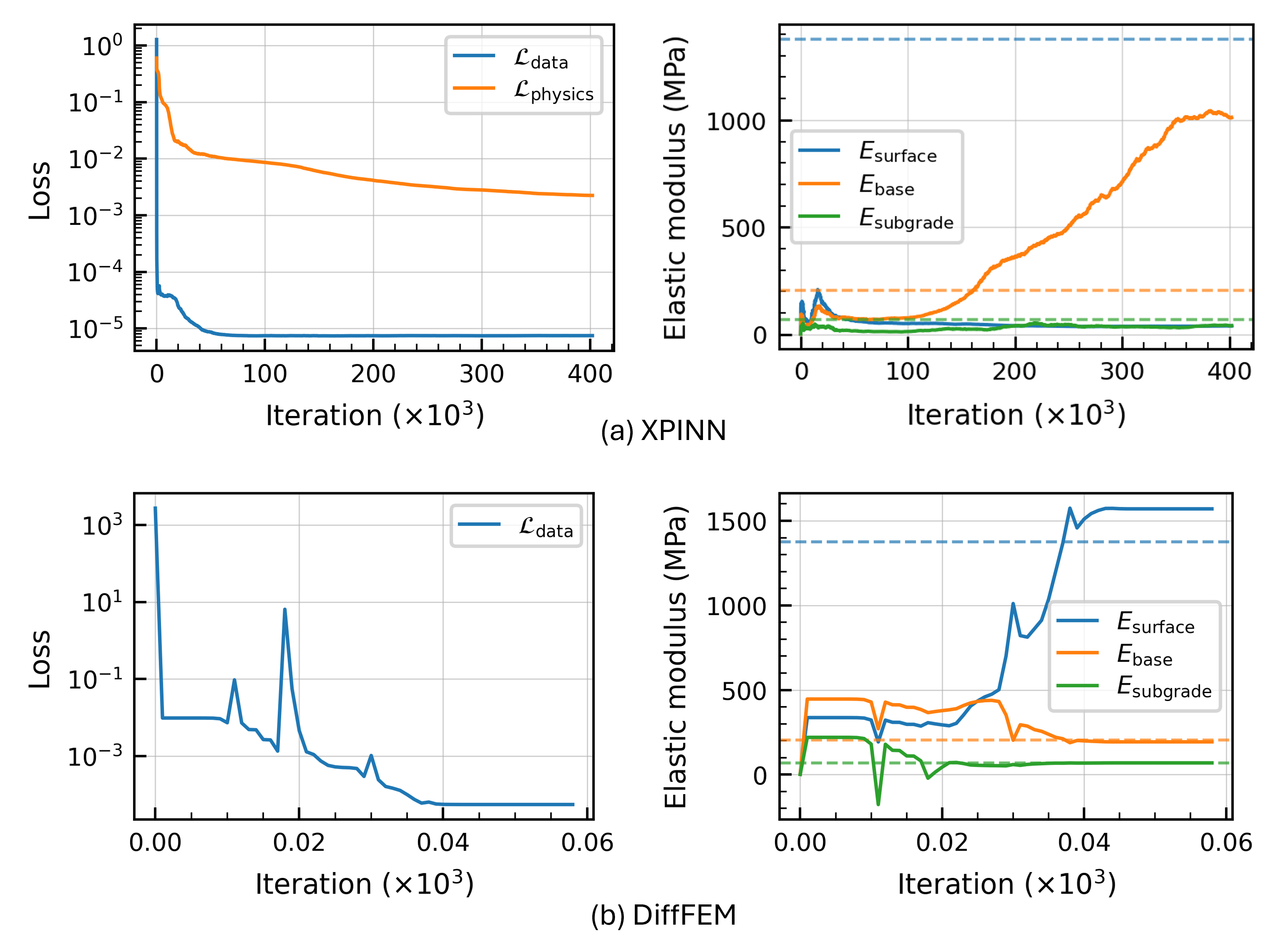}
    \caption{Convergence histories for the noisy inversion with $\sigma_n = 5\,\mu\mathrm{m}$.
    Top row: XPINN loss components and inferred moduli.
    Bottom row: DiffFEM loss and inferred moduli.
    Dashed horizontal lines indicate the ground-truth moduli.}
    \label{fig:inv_hist_noise_5}
\end{figure}

\begin{figure}[!htbp]
    \centering
    \includegraphics[width=1.0\textwidth]{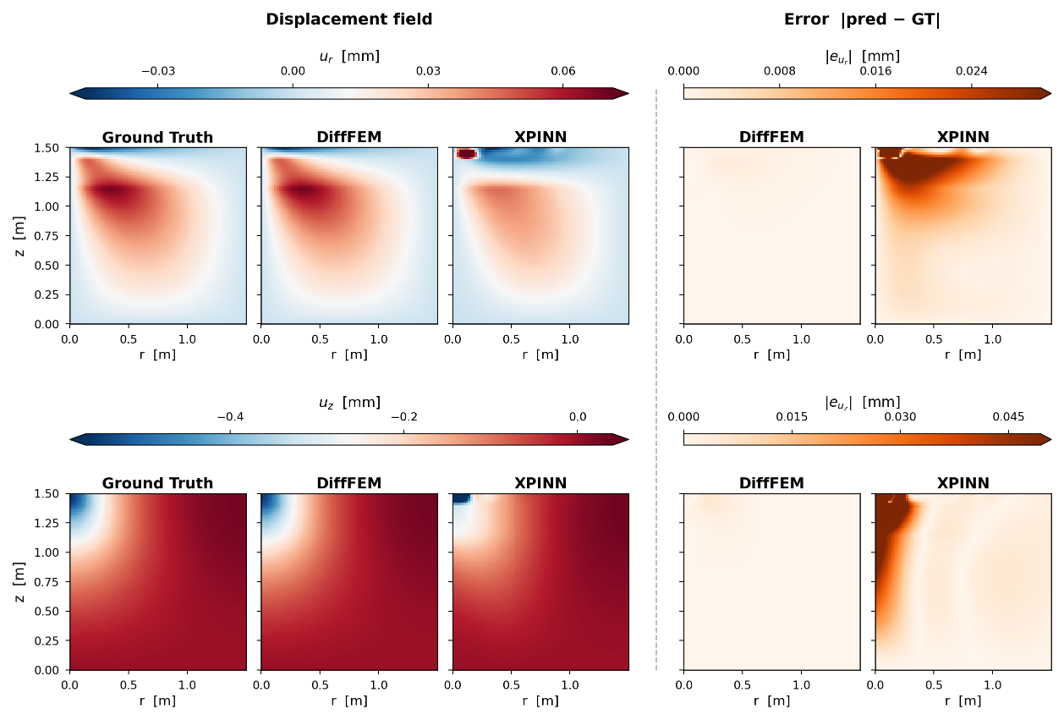}
    \caption{Predicted displacement fields and pointwise errors for $\sigma_n = 5\,\mu\mathrm{m}$.
    The plot layout follows that of \Cref{fig:displacement_comparison_noise_0}.}
    \label{fig:displacement_comparison_noise_5}
\end{figure}

Under high noise, DiffFEM (\Cref{fig:inv_hist_noise_5}b) still shows fast convergence, with the data loss plateau rising to $\sim 10^{-3}$.
The recovered moduli drift from the ground truth (14\%, 6.2\%, and 0.09\% for the surface, base, and subgrade) due to the measurement noise, but the result remains reasonable.
XPINN (\Cref{fig:inv_hist_noise_5}a) fails to identify the moduli, more severely than the case with $\sigma_n = 1~\mu$m.
The base modulus drifts monotonically upward throughout training and reaches roughly 1,000~MPa, nearly five times the true values, while the surface and subgrade estimates remain well below their true values shown in \Cref{tab:layers}.
\Cref{fig:displacement_comparison_noise_5} illustrates the consequence for the reconstructed displacement field.
DiffFEM tracks the ground truth with small errors throughout the domain for both $u_r$ and $u_z$.
The XPINN's displacement fields show a physically incorrect pattern near the loaded zone and also along the centerline below the base layer.

\subsubsection{Summary across noise levels}
\label{subsec:noise_summary}

\begin{figure}[!htbp]
    \centering
    \includegraphics[width=1.0\textwidth]{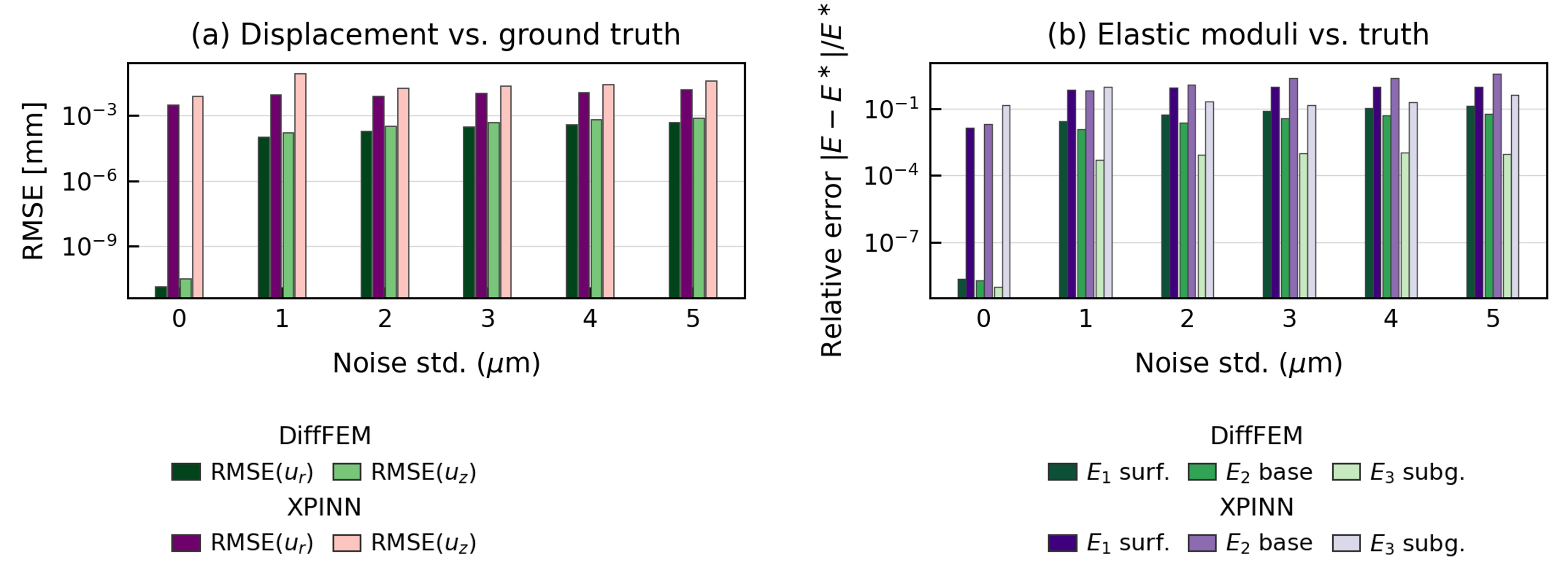}
    \caption{Summary of inversion accuracy across all noise levels ($\sigma_n = 0$--$5\,\mu\mathrm{m}$).
    (a)~Root-mean-square error of the predicted surface displacements relative to the ground-truth solution.
    (b)~Relative error of the inferred elastic moduli for each layer.
    All axes are logarithmic.}
    \label{fig:inv_result_summary}
\end{figure}

\Cref{fig:inv_result_summary} summarizes the performance of both methods across all tested noise levels, with quantitative values reported in \Cref{tab:appndx_inversion_summary}.
DiffFEM degrades gracefully with increasing noise: the displacement RMSE grows approximately linearly with $\sigma_n$, and the modulus errors remain bounded, reaching 14.0\% (surface), 6.2\% (base), and 0.09\% (subgrade) at $\sigma_n = 5~\mu$m.
XPINN, in contrast, exhibits a displacement RMSE about an order of magnitude larger than that of DiffFEM, indicating that XPINN's accuracy is limited by optimization-related errors rather than by measurement noise.
Its modulus errors are correspondingly large and erratic for $\sigma_n \geq 1~\mu$m, with base errors reaching up to 391\% and the dominant failure mode shifting with noise level---from surface-base collapse at low noise to severe base overestimation at high noise.

\subsection{\added{Robustness and extensions}}
\label{subsec:robustness}

\subsubsection{\added{Effect of model discrepancy}}
\label{subsec:model_discrepancy}

\added{The results in \Cref{subsec:noise_free,subsec:noisy} use data generated by the inversion model itself, which we refer to as the consistent-model condition.}
\added{This condition can overestimate the inversion accuracy (\Cref{subsec:benchmark}), so we test whether the DiffFEM accuracy holds when a different model generates the data.}
\added{We invert the refined-mesh data with the original mesh (\Cref{tab:model_discrepancy}).}
\added{Without noise, the model discrepancy yields errors of $-0.29\%$, $-0.31\%$, and $+0.09\%$ for the surface, base, and subgrade moduli, respectively, compared with $\mathcal{O}(10^{-9})$ under the consistent-model condition.}
\added{Data from an even finer $4\times4$ mesh change these errors by less than 0.1 percentage points, so they reflect the discretization error of the original mesh rather than the inversion.}
\added{Under noise, the errors differ from those with the original data by at most 0.4 percentage points, which is small relative to the noise-induced errors in the surface and base moduli.}

\added{The near-exact noise-free accuracy therefore relies on the consistent-model condition, but a model discrepancy has little effect on the DiffFEM accuracy under noise.}
\added{XPINN also benefits from this condition because its physics loss uses the same stiffness matrix (\Cref{eq:loss_physics}), so its noise-free errors reflect the inversion algorithm rather than a model discrepancy.}

\subsubsection{\added{Effect of positivity constraint}}
\label{subsec:positivity}

\added{
The inversions in \Cref{subsec:noise_free,subsec:noisy} do not constrain the sign of the moduli, so the optimizer can reach physically inadmissible negative values, as in the XPINN case with $\lambda_\mathrm{data}/\lambda_\mathrm{phys}=1$ (\Cref{tab:xpinn_loss_weight_sweep}). 
It is then unclear whether the XPINN failures stem from the PINN formulation or from the unconstrained parameter space. 
To separate the two, we enforce positivity by parameterizing each modulus as $E_i = \exp(\theta_i)$ and optimizing $\theta_i$ instead of $E_i$, similar to the exponential parameterization of \citet{chen2025comprehensive}. 
All other settings, including the initial moduli of 1~MPa and the XPINN budget of $4\times10^{5}$ iterations, are unchanged. 
\Cref{tab:positivity} compares the resulting errors with those of the unconstrained inversions.
}

\added{
For DiffFEM, the constraint does not change the solution.
The recovered moduli agree with the unconstrained results within $10^{-5}\%$ at all noise levels.
The optimization no longer evaluates negative moduli and converges in 35--37 iterations instead of 44--62 (\Cref{fig:inv_hist_positivity}).
For XPINN, the constraint removes the negative moduli but does not resolve the inversion failure. 
The errors remain large under noise and with $\lambda_\mathrm{data}/\lambda_\mathrm{phys}=1$, and the subgrade modulus stays near its initial value in all cases (\Cref{fig:inv_hist_positivity}), with errors between $-92\%$ and $-99\%$. 
The noise-free case also loses the accuracy obtained without the constraint.
Under the exponential parameterization, $\partial\mathcal{L}/\partial\theta_i = E_i\,\partial\mathcal{L}/\partial E_i$, so the moduli are updated slowly while they remain near the low initial value, relative to the network weights optimized jointly with them.
DiffFEM is not affected because it optimizes only the three moduli. 
Allowing negative moduli was therefore not the sole cause of the XPINN failures under the tested settings, and XPINN is also sensitive to how the moduli are parameterized.
}

\begin{table}[t]
\centering
\caption{\added{Signed relative errors (\%) of the moduli recovered without and with the positivity constraint $E_i = \exp(\theta_i)$.
XPINN uses $\lambda_\mathrm{data}/\lambda_\mathrm{phys}=10^{3}$ unless noted.
DiffFEM errors are identical without and with the constraint.}}
\label{tab:positivity}
\resizebox{\textwidth}{!}{%
    \begin{tabular}{c rrr rrr rrr}
        \toprule
        & \multicolumn{3}{c}{XPINN, unconstrained} & \multicolumn{3}{c}{XPINN, constrained} & \multicolumn{3}{c}{DiffFEM, constrained} \\
        \cmidrule(lr){2-4} \cmidrule(lr){5-7} \cmidrule(lr){8-10}
        $\sigma_n$ ($\mu$m) & $E_\mathrm{surface}$ & $E_\mathrm{base}$ & $E_\mathrm{subgrade}$ & $E_\mathrm{surface}$ & $E_\mathrm{base}$ & $E_\mathrm{subgrade}$ & $E_\mathrm{surface}$ & $E_\mathrm{base}$ & $E_\mathrm{subgrade}$ \\
        \midrule
        0 & -1.5 & +2.1 & -14.4 & -76.2 & +66.0 & -92.1 & 0.00 & 0.00 & 0.00 \\
        1 & -74.4 & +66.7 & -94.2 & -9.1 & +12.7 & -97.3 & +2.71 & -1.23 & -0.05 \\
        3 & -95.6 & +233.6 & -14.7 & -92.1 & +116.5 & -93.8 & +8.26 & -3.69 & -0.10 \\
        5 & -97.2 & +391.2 & -42.9 & -96.5 & +264.8 & -93.5 & +13.97 & -6.15 & -0.09 \\
        \midrule
        0 ($\lambda_\mathrm{data}/\lambda_\mathrm{phys}{=}1$) & -313.1 & -121.5 & -102.2 & -61.0 & +88.0 & -98.8 & -- & -- & -- \\
        \bottomrule
    \end{tabular}%
}
\end{table}

\subsubsection{\added{Scalability of DiffFEM to more layers}}
\label{subsec:four_layer}

\added{
Pavements often include more than three layers, for example a granular subbase between the base and subgrade.
To examine whether DiffFEM extends to such systems, we insert a 303~mm subbase with a typical granular modulus of 103~MPa \citep{aashto1993guide} into the upper part of the original subgrade (\Cref{tab:four_layer}).
The mesh is unchanged, and only the element grouping in \Cref{eq:stiffness_decomp} changes.
We invert the four moduli using DiffFEM with the previous analysis setting.
}

\added{
Without noise, DiffFEM recovers all four moduli with errors below $10^{-6}\%$ in 46 iterations, similar to the three-layer case, and the cost per iteration is unchanged.
Under noise, DiffFEM converges in 41--50 iterations (\Cref{fig:inv_hist_four_layer}), and the errors grow approximately linearly with $\sigma_n$, reaching 19\% for the surface modulus and 6\% for the subgrade modulus at $\sigma_n = 5~\mu\mathrm{m}$.
These errors exceed those of the three-layer case (\Cref{tab:appndx_inversion_summary}) because the subbase and subgrade affect the surface deflections in a similar way, so an increase in one modulus can be largely offset by a decrease in the other.
Since the noise-free inversion is exact, the larger errors reflect the weaker constraint that the data place on four moduli, not the inversion algorithm.
}

\begin{table}[t]
\centering
\caption{\added{Four-layer configuration and signed relative errors (\%) of the moduli recovered by DiffFEM.}}
\label{tab:four_layer}
    \setlength{\tabcolsep}{5pt}
    \begin{tabular}{lccc rrrrrr}
        \toprule
        & & & & \multicolumn{6}{c}{Error (\%) at $\sigma_n$ ($\mu$m)} \\
        \cmidrule(lr){5-10}
        Layer & Thickness & $E$ (MPa) & $\nu$ & 0 & 1 & 2 & 3 & 4 & 5 \\
        \midrule
        Surface (AC) & 100\,mm & 1{,}378 & 0.35 & 0.00 & +3.53 & +7.20 & +11.00 & +14.94 & +19.03 \\
        Base course & 250\,mm & 206 & 0.40 & 0.00 & -2.15 & -4.35 & -6.60 & -8.90 & -11.25 \\
        Subbase & 303\,mm & 103 & 0.40 & 0.00 & +1.75 & +3.67 & +5.78 & +8.09 & +10.61 \\
        Subgrade & 847\,mm & 69 & 0.45 & 0.00 & -1.27 & -2.54 & -3.80 & -5.04 & -6.27 \\
        \bottomrule
    \end{tabular}
\end{table}

\subsubsection{\added{Comparison with finite-difference gradients}}
\label{subsec:fd_baseline}

\added{
DiffFEM obtains the exact gradient through AD (\Cref{subsec:ad}).
To compare DiffFEM with the conventional FD approach (\Cref{subsec:fd}), we repeat the four-layer inversion (\Cref{subsec:four_layer}) with the gradient replaced by the forward-difference estimate in \Cref{eq:fd}.
We use the same L-BFGS optimizer, initial moduli, and data, so only the gradient differs.
The perturbation is set relative to each modulus, $\Delta E_i = hE_i$, with $h=10^{-8}$, $10^{-2}$, and $10^{-1}$.
Each iteration then requires $n+1=5$ forward solves, whereas DiffFEM requires one due to AD capacity.
}

\added{
\Cref{tab:fd_baseline} compares the noise-free inversions, in which the errors reflect the gradient alone, for representative perturbations.
With $h=10^{-8}$, the FD inversion recovers the true moduli as DiffFEM does, but it needs five times as many forward solves per iteration.
For example, both converge in 46 iterations, which amounts to 230 forward solves for FD and 46 for DiffFEM.
As $h$ increases further, the gradient error stops the inversion at incorrect moduli, with errors of up to 12\% at $h=10^{-2}$ and 30\% at $h=10^{-1}$.
The FD approach therefore matches DiffFEM only when $h$ is tuned, whereas the AD gradient requires no such choice.
Because the FD cost per iteration grows with the number of unknowns while the AD gradient can be obtained from one forward solve, the computational advantage of DiffFEM increases for more layers, and this advantage becomes significant for spatially varying properties (e.g., soil parameter field).
}

\begin{table}[t]
\centering
\caption{\added{Noise-free four-layer inversion with the AD gradient (DiffFEM) and with forward-difference (FD) gradients of relative perturbation $h$, using the same L-BFGS optimizer.}}
\label{tab:fd_baseline}
    \setlength{\tabcolsep}{5pt}
    \sisetup{table-format=+2.2, retain-explicit-plus}
    \begin{tabular}{lcccSSSS}
        \toprule
        \multirow{2}{*}{Gradient} & \multirow{2}{*}{$h$} & Forward solves & \multirow{2}{*}{Iterations} & \multicolumn{4}{c}{Relative error (\%)} \\
        \cmidrule(lr){5-8}
        & & per iteration & & {$E_\mathrm{surface}$} & {$E_\mathrm{base}$} & {$E_\mathrm{subbase}$} & {$E_\mathrm{subgrade}$} \\
        \midrule
        AD (DiffFEM) & -- & 1 & 46 & 0.00 & 0.00 & 0.00 & 0.00 \\
        \midrule
        FD & $10^{-8}$ & 5 & 46 & 0.00 & 0.00 & 0.00 & 0.00 \\
        FD & $10^{-2}$ & 5 & 86 & +1.08 & +2.39 & -12.18 & +11.96 \\
        FD & $10^{-1}$ & 5 & 58 & +29.98 & -14.74 & -5.41 & +10.81 \\
        \bottomrule
    \end{tabular}
\end{table}

\subsection{Computational cost}

We compare the computational efficiency of DiffFEM and XPINN on the noise-free inversion shown in \Cref{fig:inv_hist_noise_0}, using a NVIDIA RTX 5000 Ada GPU.
\added{Both methods run in double precision and use the PyTorch L-BFGS optimizer with a history size of 200 and a strong Wolfe line search.}
\added{DiffFEM solves \Cref{eq:ksystem} with a dense LU factorization.}
\added{The full implementation details are available in the open-source code (Section~\ref{sec:code_availability}).}
DiffFEM completes the full optimization iterations in approximately 60~s, whereas XPINN requires roughly 10~h to reach an approximate stationary point after $400 \times 10^{3}$ iterations---more than two orders of magnitude slower.
\added{This difference stems mainly from the number of iterations, since DiffFEM converges in only 44 iterations.}

\section{Discussion}

The results above consistently show that DiffFEM outperforms XPINN in accuracy, robustness, and computational efficiency for FWD backcalculation.
This observation raises a practical question: why does DiffFEM perform reliably in this setting, and under what conditions should each approach be preferred?

The superior performance of DiffFEM in this study appears to stem from several features of its formulation.
First, the governing equations are enforced directly through the forward solve at every iteration, whereas in XPINN, they are imposed only as soft constraints in the loss function and therefore need not be satisfied exactly.
Second, DiffFEM optimizes only the three layer moduli, while XPINN must simultaneously optimize the moduli, tens of thousands of neural-network weights, and interface-continuity terms.
In addition, the PINN loss is a composite objective, requiring the optimization process to balance multiple competing terms.
This can make the loss landscape more complex and increase the likelihood of convergence to local optima \citep{rowan2026visualizing}.
For this reason, previous studies have often adopted problem-specific architectural modifications and strategic adaptive optimization policy with second-order methods \citep{rathore2024challenges,moon2025physics}.
In the present setting, these factors make the PINN-based approaches challenging to train and can hinder reliable parameter identification in the presence of noisy data.

The practical advantage of DiffFEM, however, rests on an important prerequisite: the forward solver must be differentiable.
In the present study, this requirement is straightforward to satisfy.
Layered elastic theory under quasi-static loading reduces to a linear system (\Cref{eq:ksystem}) that is readily implemented within modern AD frameworks such as PyTorch.
There are no adaptive mesh updates, and the constitutive model is linear elastic---features that simplify the use of AD in this setting.
Many layered-elastic FWD backcalculation problems share these characteristics, which suggests that DiffFEM can be a strong option for this class of problems.

Not all forward solvers, however, can be made differentiable with equal ease.
Contact algorithms involving non-smooth constraints, constitutive stress update with adaptive timestepping, and bifurcation problems with multiple solution branches can all present obstacles to AD-based differentiation: the gradient of the forward solution with respect to the material parameters can be undefined, discontinuous, or expensive to propagate through the computational graph.
Making such solvers differentiable is an active area of research \citep{paulus2026softjax}, and in settings where a differentiable forward solver is difficult to construct, PINN-based approaches remain an attractive option.
For example, PINNs have shown successful applications for contact problems \citep{sahin2024solving, bai2025energy, li2024physics} and elastoplastic constitutive update with return mapping \citep{eghbalian2023physics}.
In addition, PINNs can avoid an explicit linear solve and the need for a conforming finite element mesh in their classical form, where PINNs evaluate the particle differential equation residual by applying AD to the network output with respect to the input coordinates \citep{raissi2019physics}.
Therefore, observations at arbitrary sensor locations can be incorporated directly, while complex geometries or regions requiring higher local resolution can be addressed through targeted collocation-point sampling, although boundary-condition enforcement and sampling design remain important challenges \citep{cuomo2022scientific}.

We note, however, that the XPINN variant benchmarked in this study is not mesh-free in this sense.
Its physics residual is assembled from precomputed FEM stiffness matrices (\Cref{eq:stiffness_decomp}), so it inherits the mesh-generation and Degree of Freedom indexing machinery of DiffFEM while replacing only the linear solve with a neural parameterization of the displacement field.
The mesh-free advantages above therefore apply to autograd particle differential equation residual PINNs \citep{raissi2019physics}, not to the specific XPINN implementation compared here.

Beyond the residual formulation, the XPINN domain-decomposition strategy introduces its own scaling concern: each additional pavement layer requires a dedicated sub-network, and each interior interface contributes a continuity penalty, so that both the total parameter count and the number of interface-matching terms grow linearly with the number of layers.
DiffFEM, by contrast, scales more favorably, since adding a layer only introduces one additional unit stiffness matrix in the additive decomposition (\Cref{eq:stiffness_decomp}) and one additional scalar modulus in the optimization, without altering the structure of the inverse problem or the linear system being solved.
\added{The four-layer inversion in \Cref{subsec:four_layer} supports this: adding a subbase changes only the element grouping, and the noise-free inversion remains exact with a similar number of iterations.}

A further consideration in DiffFEM is the memory footprint of reverse-mode AD.
For problems involving long time integration, such as dynamic FWD analysis or viscoelastic pavement response, storing intermediate states for backpropagation can become memory-prohibitive.
To mitigate this issue, DiffFEM may require specialized techniques such as gradient checkpointing \citep{choi2024inverse} to efficiently manage the memory.
These complications do not arise in the quasi-static setting considered here.
In such time-dependent extensions, PINNs are an appealing option, as they avoid unrolling a time-stepping computational graph.
For instance, \citet{hou2024physics} solved long-horizon forward and inverse problems efficiently without timestepping on wave propagation based on PINN.

Taken together, these observations suggest a practical guideline.
When the forward solver is well established, numerically stable, and amenable to implementation within an AD framework---as in the layered-elastic FWD backcalculation problem studied here---DiffFEM appears to be a good starting option.
In this setting, it imposes hard physics constraints that enforce equilibrium at every iteration through the forward solve, lets the optimization focus on system parameters without considering neural network surrogates, and converges in minutes rather than hours in the present benchmark.
PINNs, while conceptually appealing for their ability to simultaneously solve the forward and inverse problems through soft physics constraints, should be applied with caution to problems of this type: the results of this study show that even with domain decomposition and careful hyperparameter tuning, the PINN formulation struggled to recover accurate modulus estimates under realistic noise conditions.
PINNs may be more compelling in settings where FEM faces computational or numerical challenges in solving the governing equations, or when a differentiable forward solver is unavailable.

\section{Conclusion}

This study investigates recent AD-based inverse analysis approaches, PINN and DiffFEM, for backanalysis of pavement modulus from FWD tests.
The results show that the suitability of these two approaches differs fundamentally for FWD backanalysis.

The standard PINN formulation does not recover the layer moduli due to the sharp material discontinuities inherent to layered pavements.
Although XPINN with domain decomposition improves convergence in the noise-free case, its performance remains highly sensitive to the hyperparameter selection, including loss-weight selection and network architecture.
Even after hyperparameter tuning, XPINN requires substantially more iterations and still produces suboptimal solutions because of the inherent formulation of PINNs, where governing equations and interface continuity are imposed only as soft constraints.

On the other hand, DiffFEM provides consistently more accurate and robust inversion results than XPINN.
In the noise-free case, it recovers the target moduli to numerical precision.
Under all tested noise levels, it remained stable and degraded gradually with increasing noise, while maintaining substantially lower displacement and modulus errors than XPINN.
This better performance can be because DiffFEM enforces the governing equations exactly at every iteration as a hard constraint through the FEM forward solver integrated with the AD.
In addition, DiffFEM is far more computationally efficient.
DiffFEM converges in about 60~s, whereas XPINN required about 10~h on the same hardware.

Overall, the findings indicate that PINN-based inverse analysis should be used with caution for FWD backcalculation.
For problems like layered-elastic systems, where the forward model is well established, numerically stable, and differentiable, DiffFEM is a more reliable and efficient choice.
More broadly, this study suggests that in inverse problems with accessible differentiable solvers, physics-native AD formulations may offer a more practical path than PINN-based approaches, where physics-consistent solutions are not always guaranteed.
Nevertheless, PINNs may remain attractive when differentiable forward solvers are unavailable, difficult to implement, computationally burdensome due to long time-stepping histories, or when mesh-free formulations offer practical advantages.

A potential direction for future work is to evaluate the framework under more realistic pavement conditions, including viscoelasticity and nonlinear constitutive behavior.
These settings may introduce adaptive time stepping and non-differentiable algorithmic components, which could complicate differentiable solver construction.
Whether we can make these operations differentiable to take advantage of AD remains unclear and opens opportunities for further studies.
\added{Another direction is to validate the framework with field FWD data.}

\section{Data and code availability}
\label{sec:code_availability}
\added{The source code of the DiffFEM and PINN implementations, the synthetic datasets, and the trained network weights are available at \url{https://github.com/yjchoi1/diff-pave-paper}.}

\section{Acknowledgment}
This work was supported by the InnoCORE program of the Ministry of Science and ICT(N10260002). 
The authors also acknowledge support from the University of Massachusetts Amherst Riccio College of Engineering Startup Fund for new faculty.

\clearpage
\renewcommand{\theequation}{\Alph{section}.\arabic{equation}}
\renewcommand{\thefigure}{\Alph{section}.\arabic{figure}}
\renewcommand{\thetable}{\Alph{section}.\arabic{table}}
\appendix
\setcounter{table}{0}
\setcounter{figure}{0}

\section{}
\label{app:appendix_a}

\begin{table}[!h]
  \centering
    \caption{Quantitative inversion results across all noise levels. 
    True moduli: $E_{surface} = 1{,}378$~MPa, $E_{base} = 206$~MPa, 
    $E_{subgrade} = 69$~MPa.
    This table is visualized as \Cref{fig:inv_result_summary}}
    \label{tab:appndx_inversion_summary}
      \resizebox{\textwidth}{!}{%
  \footnotesize
  \setlength{\tabcolsep}{4pt}
  \begin{tabular}{@{} c *{10}{r} @{}}
    \toprule
    $\sigma$ &
    \multicolumn{4}{c}{RMSE (mm)} &
    \multicolumn{6}{c}{Relative error (\%)} \\
    \cmidrule(lr){2-5} \cmidrule(lr){6-11}
    (\si{\micro\metre}) &
    \multicolumn{2}{c}{$u_r$} & \multicolumn{2}{c}{$u_z$} &
    \multicolumn{2}{c}{$E_1$ (surface)} &
    \multicolumn{2}{c}{$E_2$ (base)} &
    \multicolumn{2}{c}{$E_3$ (subgrade)} \\
    \cmidrule(lr){2-3} \cmidrule(lr){4-5}
    \cmidrule(lr){6-7} \cmidrule(lr){8-9} \cmidrule(lr){10-11}
    & DiffFEM & XPINN & DiffFEM & XPINN & DiffFEM & XPINN & DiffFEM & XPINN & DiffFEM & XPINN \\
    \midrule
    0 &
      $1.36\times10^{-11}$ & 0.00291 &
      $3.12\times10^{-11}$ & 0.00719 &
      $2.31\times10^{-7}$  & 1.46 &
      $2.01\times10^{-7}$  & 2.13 &
      $9.89\times10^{-8}$ & 14.4 \\
    1 &
      $9.54\times10^{-5}$ & 0.00892 &
      $1.55\times10^{-4}$ & 0.0835 &
      2.71 & 74.4 &
      1.23 & 66.7 &
      $5.04\times10^{-2}$ & 94.2 \\
    2 &
      $1.89\times10^{-4}$ & 0.00755 &
      $3.08\times10^{-4}$ & 0.0178 &
      5.46 & 92.5 &
      2.46 & 122 &
      $8.46\times10^{-2}$ & 21.1 \\
    3 &
      $2.80\times10^{-4}$ & 0.0100 &
      $4.59\times10^{-4}$ & 0.0224 &
      8.26 & 95.6 &
      3.69 & 234 &
      $1.03\times10^{-1}$ & 14.7 \\
    4 &
      $3.69\times10^{-4}$ & 0.0105 &
      $6.07\times10^{-4}$ & 0.0260 &
      11.1  & 96.0 &
      4.92 & 245 &
      $1.06\times10^{-1}$ & 19.2 \\
    5 &
      $4.56\times10^{-4}$ & 0.0147 &
      $7.53\times10^{-4}$ & 0.0381 &
      14.0  & 97.2 &
      6.15 & 391 &
      $9.41\times10^{-2}$ & 42.9 \\
    \bottomrule
  \end{tabular}
  }
\end{table}

\begin{table}[!h]
\centering
\caption{\added{Signed relative errors (\%) of the DiffFEM moduli inverted from the refined-mesh data and from the original data.}}
\label{tab:model_discrepancy}
    \begin{tabular}{c rrr rrr}
        \toprule
        & \multicolumn{3}{c}{Refined-mesh data} & \multicolumn{3}{c}{Original data} \\
        \cmidrule(lr){2-4} \cmidrule(lr){5-7}
        $\sigma_n$ ($\mu$m) & $E_\mathrm{surface}$ & $E_\mathrm{base}$ & $E_\mathrm{subgrade}$ & $E_\mathrm{surface}$ & $E_\mathrm{base}$ & $E_\mathrm{subgrade}$ \\
        \midrule
        0 & -0.29 & -0.31 & +0.09 & 0.00 & 0.00 & 0.00 \\
        1 & +2.41 & -1.53 & +0.04 & +2.71 & -1.23 & -0.05 \\
        2 & +5.14 & -2.76 & +0.00 & +5.46 & -2.46 & -0.09 \\
        3 & +7.92 & -3.98 & -0.02 & +8.26 & -3.69 & -0.10 \\
        4 & +10.74 & -5.20 & -0.02 & +11.09 & -4.92 & -0.11 \\
        5 & +13.60 & -6.43 & -0.01 & +13.97 & -6.15 & -0.09 \\
        \bottomrule
    \end{tabular}
\end{table}

\begin{figure}[!htbp]
    \centering
    \includegraphics[width=\textwidth]{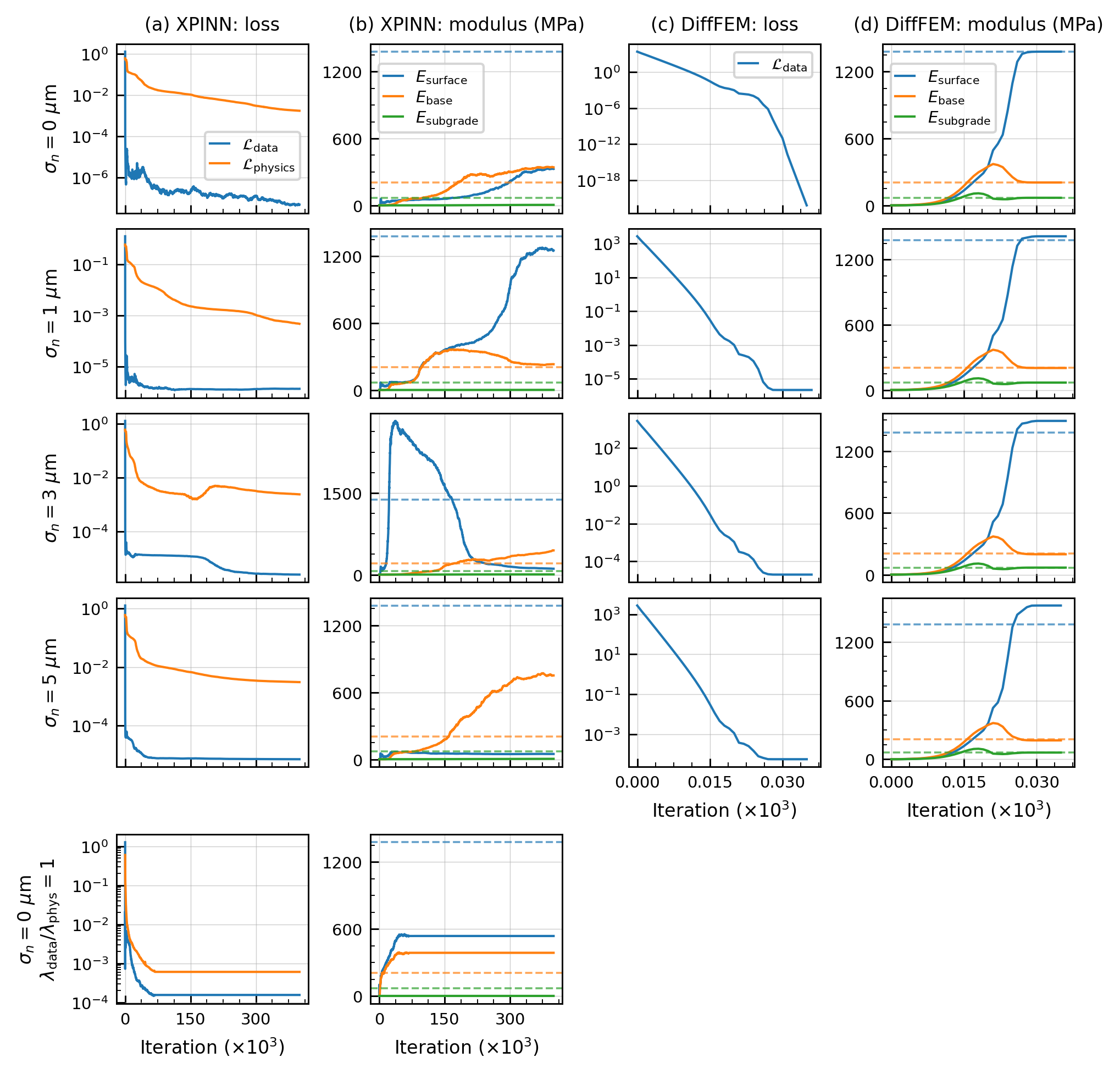}
    \caption{\added{Convergence histories of XPINN and DiffFEM with the positivity constraint $E_i = \exp(\theta_i)$.
    Columns show the loss components and inferred moduli of each method, and rows show the noise levels.
    XPINN uses $\lambda_\mathrm{data}/\lambda_\mathrm{phys}=10^{3}$ except in the bottom row.
    Dashed horizontal lines indicate the ground-truth moduli.}}
    \label{fig:inv_hist_positivity}
\end{figure}

\begin{figure}[!htbp]
    \centering
    \includegraphics[width=0.7\textwidth]{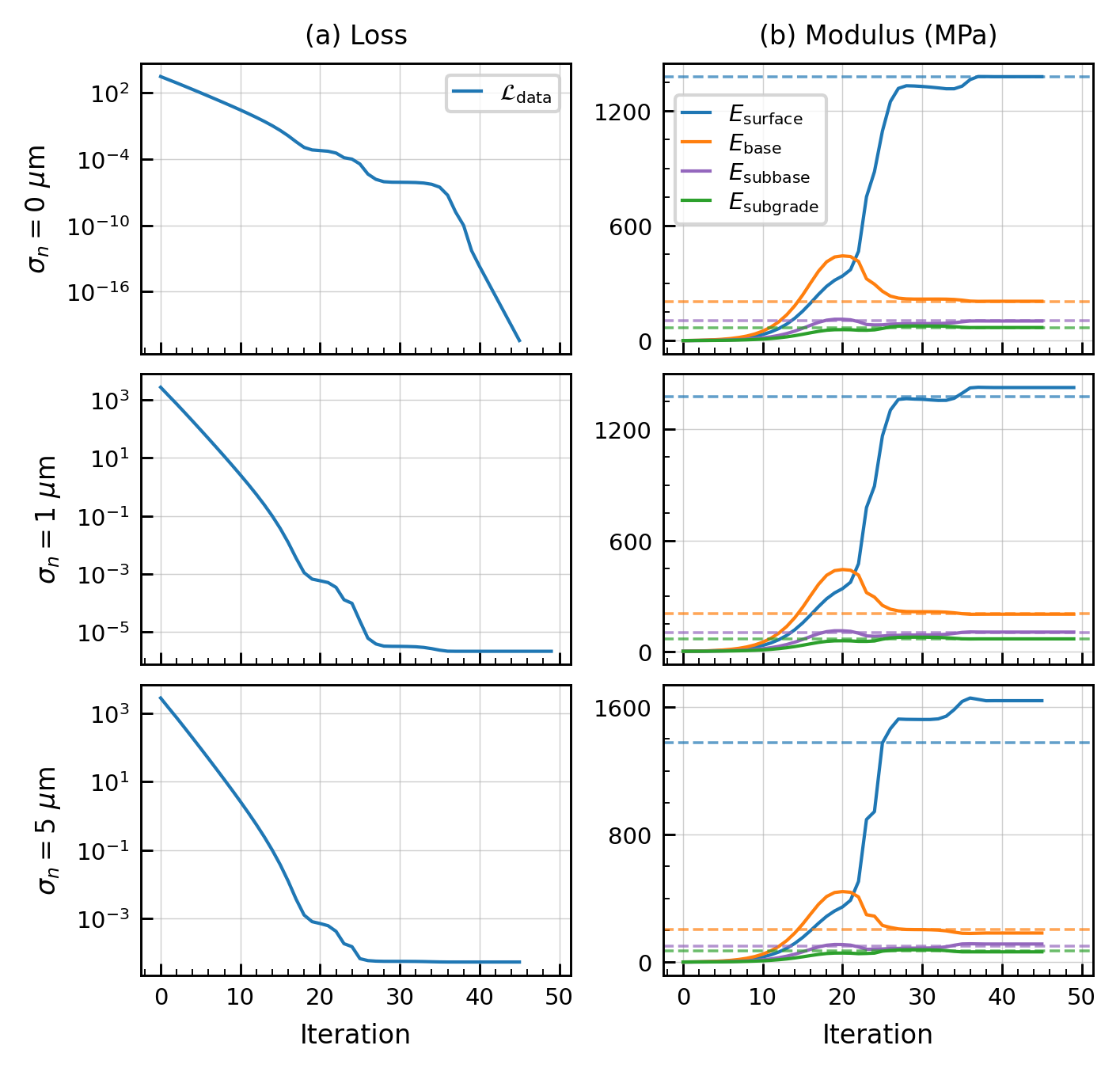}
    \caption{\added{Convergence histories of the four-layer DiffFEM inversion.
    Rows show the noise levels.
    Dashed horizontal lines indicate the ground-truth moduli.}}
    \label{fig:inv_hist_four_layer}
\end{figure}



\clearpage
\bibliographystyle{elsarticle-harv} 
\bibliography{main}






\end{document}